%% file: main.tex
\documentclass[
 amsmath,amssymb,
 aps,twocolumn,superscriptaddress
]{revtex4-2}

\usepackage{graphicx}
\usepackage{dcolumn}
\usepackage{bm}
\usepackage{hyperref}

\usepackage{cleveref}
\crefname{equation}{Eq.}{Eqs.}
\crefname{equations}{Eqs.}{Eqs.}
\Crefname{equation}{Equation}{Equations}
\crefname{figure}{Fig.}{Figs.}
\Crefname{figure}{Figure}{Figures}
\crefname{section}{Sec.}{Secs.}
\Crefname{section}{Section}{Sections}
\crefname{appendix}{Appendix}{Apps.}
\Crefname{appendix}{Appendix}{Apps.}
\crefname{paragraph}{Sec.}{Secs.}
\crefname{table}{Table}{Tables}
\crefname{algorithm}{Box}{Boxes}
\Crefname{algorithm}{Box}{Boxes}
\usepackage{physics}

\input{commands}

\begin{document}
	\title{Quantum sensing of displacements with stabilized GKP states}
	\author{Lautaro Labarca}
    \email{labl2714@usherbrooke.ca}
	\affiliation{Institut Quantique and Département de Physique, Université de Sherbrooke, Sherbrooke, Qu\'ebec J1K 2R1, Canada}
    \author{Sara Turcotte}
    \affiliation{Institut Quantique and Département de Physique, Université de Sherbrooke, Sherbrooke, Qu\'ebec J1K 2R1, Canada}
    \author{Alexandre Blais}
	\affiliation{Institut Quantique and Département de Physique, Université de Sherbrooke, Sherbrooke, Qu\'ebec J1K 2R1, Canada}
    \affiliation{Canadian Institute for Advanced Research, Toronto, ON M5G 1M1, Canada}
    \author{Baptiste Royer}
    \affiliation{Institut Quantique and Département de Physique, Université de Sherbrooke, Sherbrooke, Qu\'ebec J1K 2R1, Canada}
    
	\begin{abstract}
	We demonstrate how recent protocols developed for the stabilization of Gottesman-Kitaev-Preskill (GKP) states can be used for the estimation of two-quadrature displacement sensing, with sensitivities approaching the multivariate quantum Cramer-Rao bound. 
    Thanks to the stabilization, this sensor is backaction evading and can function continuously without reset, making it well suited for the detection of itinerant signals. Additionally, we provide numerical simulations showing that the protocol can unconditionally surpass the Gaussian limit of displacement sensing with prior information, even in the presence of realistic noise. Our work shows how reservoir engineering in bosonic systems can be leveraged for quantum metrology, with potential applications in force sensing, waveform estimation and quantum channel learning.
    \end{abstract}
    
    \maketitle
    
    \section{Introduction}

    In recent years, quantum metrology has seen rapid progress, and promises to significantly improve sensing capabilities for both applied and fundamental domains in physics, engineering and biology \cite{giovannetti2011advances,ye2024essay,aslam2023quantum}. For many of these applications multiple parameters are estimated simultaneously, such as in magnetic field sensing, force sensing, and quantum systems learning \cite{degen2017quantum,gilmore2021quantum,gebhart2023learning}. In all these applications, the limits of measurement sensitivity are set by the unavoidable noise in real experiments \cite{huelga1997improvement} and, ultimately, by the laws of quantum mechanics \cite{caves1980measurement,braunstein1994statistical}. 
    Crucially, when noise is ignored, theory predicts that quantum sensing strategies can substantially outperform classical ones \cite{bondurant1984squeezed,yurke19862, bollinger1996optimal,lee2002quantum,matsumoto2002new,giovannetti2004quantum,carollo2019quantumness}.
    When noise is considered, however, the advantage gained by the quantum strategies over classical ones can be lost and is only preserved under specific constraints, either of time, number of probes, or type of noise \cite{andre2004stability,ji2008parameter,demkowicz2009quantum,kolodynski2010phase,kacprowicz2010experimental,escher2011general,demkowicz2012elusive,hosten2016measurement}.
    
    A pathway to overcome these detrimental effects due to noise in quantum sensors, stems from the fact that quantum metrology experiments are fundamentally quantum computations \cite{aharonov2022quantum}. As such, quantum error correction (QEC) can enhance the achievable accuracy of metrology protocols in the presence of noise, something pointed out in Refs.~\cite{preskill2000quantumclocksynchronizationquantum,lee2002quantum}, and further developed in Refs.~\cite{ozeri2013heisenberglimitedmetrologyusing,dur2014improved,arrad2014increasing,kessler2014quantum,zhou2021error}. 
    In parallel, quantum information processing with bosonic systems \cite{chuang1997bosonic,cochrane1999macroscopically,gottesman2001encoding} has witnessed significant advances in recent years. Universal control of bosonic modes has been investigated theoretically and demonstrated experimentally \cite{krastanov2015universal, heeres2017implementing,Eickbusch:2022}, leading to quantum error correction beyond break-even with the Gottesman-Kitaev-Preskill (GKP) encoding~\cite{gottesman2001encoding} with superconducting circuits~\cite{Sivak:2023,lachance2024autonomous}. Successful preparation and stabilization of GKP states have also been demonstrated in trapped ions \cite{fluhmann2019encoding,DeNeeve:2022}, with ongoing efforts in optical platforms \cite{fabre2020generation,bourassa2021blueprint,konno2024logical}.
    
    In this article, we leverage recent developments in quantum engineering of bosonic systems with GKP encodings \cite{fluhmann2019encoding,DeNeeve:2022,Royer:2020,campagne2020quantum,Sivak:2023,lachance2024autonomous,zheng2024performance} to design a backaction evading sensor that estimates in a single-shot an unknown displacement in both quadratures of the bosonic system~\cite{genoni2013optimal,steinlechner2013quantum,bradshaw2018ultimate,caves2020reframing,zander2021full,Duivenvoorden:2017,Hanamura:2021,hanamura2023single,frigerio2025joint}.  
    A backaction evading sensor is one where the measurement itself prepares the sensor for the next round of estimation. Accurate backaction evading single-shot estimation is critical in the sensing of itinerant signals, where constant monitoring is needed \cite{caves2020reframing}. This is the case, for example, in gravitational wave sensing \cite{aasi2015advanced}, electric field sensing \cite{gilmore2021quantum}, and quantum illumination \cite{lloyd2008enhanced}.

    \begin{figure*}[t]
        \centering
        \includegraphics[width=\linewidth]{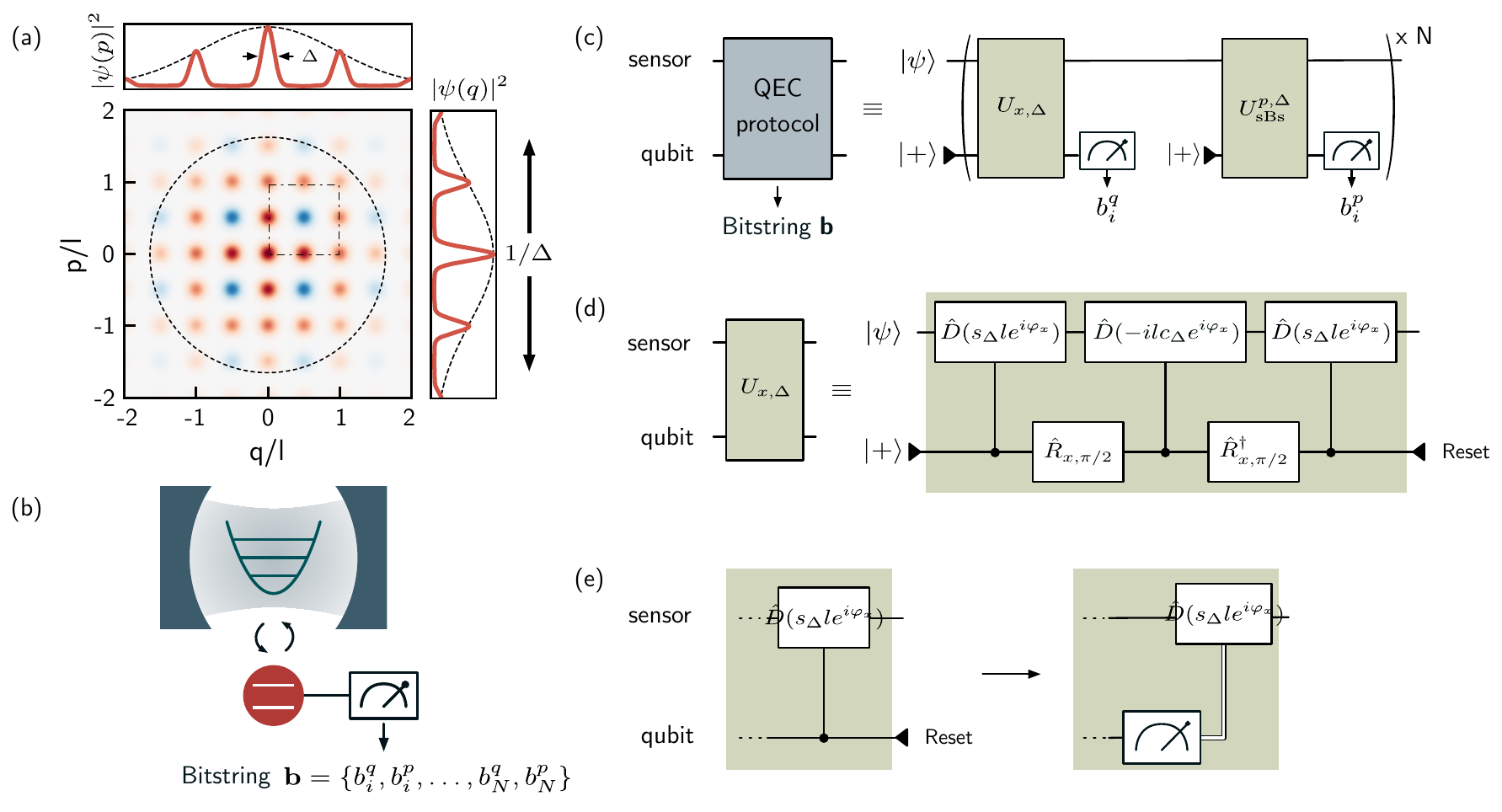}
        \caption{a) Wigner function and marginals of the finite-energy qunaught GKP state. b) Qubit-cavity pair. By entangling the qubit with a cavity mode information of the cavity is extracted via qubit measurements. c) sBs protocol. Two unitaries between the qubit and cavity are intertwined by qubit reset and repeated. Each unitary is associated to each one of the quadratures. d)sBs unitaries. Each one of the unitaries is decomposed in three control displacements, intertwined with qubit rotations.  e) For metrology, the reset of the qubit is substituted by measurement and feedback.}
        \label{fig:setup}
    \end{figure*}

    Here, we propose a protocol for two-quadrature displacement estimation that uses GKP states as sensor states, which nearly saturates the single-mode quantum bound of two-quadrature displacement sensing. This protocol does not rely on postselection or entanglement, while being backaction evading, and robust against decoherence. Additionally, it can surpass the Gaussian limit of sensing \cite{Hanamura:2021, hanamura2023single}, which is stricter than the standard quantum limit. We note that quantum advantage over the shot-noise limit using GKP states was demonstrated experimentally in Ref.~\cite{valahu2024quantum}.
     That work focused on improving the precision scaling with the total number of shots, while here we focus on improving the sensitivity of each shot. As such, our work and that of Ref.~\cite{valahu2024quantum} naturally complement each other. Given that displacement sensing is a primitive task in force sensing, waveform estimation, and quantum channel learning, our work is a step forward in the development of general-purpose bosonic quantum sensors robust against decoherence.

    The article is organized as follows. 
    In \cref{sec: Preliminaries} we provide necessary theory preliminaries on the GKP code. In \cref{sec:description-protocol}, \cref{sec:metrological-potential} and \cref{sec:backaction} we present our findings in the absence of noise. Concretely, in \cref{sec:description-protocol} we describe in detail the metrology protocol. In \cref{sec:metrological-potential} we discuss our chosen measures of metrological potential, and the performance achieved. In \cref{sec:backaction}, we study the backaction evading performance of the sensor. Moving on from the noiseless case, in \cref{sec:noise} we carry out numerical simulations showing the robustness against decoherence. In \cref{sec:discussion} we provide conclusions and outlook on future venues of research. Additionally, appendices cover accompanying results and detailed descriptions of our methods. This includes a background review on displacement sensing and QEC assisted metrology, as well as a simple derivation of the multivariate quantum Cramer-Rao bound, and an upper bound on the Holevo bound for two-quadrature displacement estimation. 

    \section{Preliminaries}
    \label{sec: Preliminaries}

    The GKP or grid code is a stabilizer bosonic code designed to correct displacement errors \cite{gottesman2001encoding}. It has been shown to saturate the capacity of the displacement channel, and to achieve near-optimal capacity when used over loss and amplification channels \cite{harrington2001achievable,zheng2024performance}. These theoretical performance metrics translate into robustness to decoherence. This robustness has been demonstrated by the successful preparation and stabilization of these states in trapped ions \cite{fluhmann2019encoding,DeNeeve:2022,valahu2024quantum}, and superconducting circuits \cite{campagne2020quantum,Sivak:2023,lachance2024autonomous}.
    
    In its simplest version, the GKP code consists of only one state, referred to as the qunaught state, and doesn't encode any logical information \cite{Duivenvoorden:2017}. It forms a square grid in phase space with lattice spacing $l = \sqrt{2\pi}$, see \cref{fig:setup} a). The ideal square qunaught state has two commuting stabilizers, $\hat T_q = e^{il\hat q}$ and $\hat T_p = e^{-il\hat p}$. 
    The infinite-energy qunaught state $\ket{\#_0}$ is the unique joint eigenstate of the stabilizers $\hat T_{x=q,p} \ket{\#_0} = \ket{\#_0}$. The qunaught state is, however, non-physical as it has infinite energy. The finite-energy qunaught state is defined as $\ket{\#_\Delta} = \hat E_\Delta \ket{\#_0}$, with  $\hat E_\Delta = e^{-\Delta^2 \hat n}$ a Gaussian envelope. This state has finite energy stabilizers $\hat T_{x,\Delta} = \hat E_\Delta \hat T_x \hat E_\Delta^{-1}$, and mean photon number $n_\Delta \simeq (1-\Delta^2)/2\Delta^2$. We refer the reader to Refs.~\cite{Matsuura:2020,Mensen-Baragiola-Menicucci:2021} for explicit expressions of its wavefunctions and Wigner representations, and to Ref.~\cite{brady2024advances} for a recent review. 

    For the preparation and stabilization of GKP states in the laboratory, consider the situation depicted in \cref{fig:setup} b), where an auxiliary qubit is coupled to a bosonic mode. This can be realized for example in superconducting circuits, where a transmon is placed inside a microwave cavity \cite{Sivak:2023,lachance2024autonomous}, or in trapped ions where the internal spin degrees of freedom couple to the mechanical motion of the ion \cite{fluhmann2019encoding,DeNeeve:2022}. Assuming the auxiliary qubit can be controlled with high fidelity, and controlled displacements can be engineered between the qubit and the bosonic mode (henceforth cavity), then GKP states can be prepared and stabilized. A control displacement is a unitary operation, where the qubit state controls a displacement on the cavity. Explicitly, $C\hat D(\alpha) = \exp((\alpha \hat a^\dagger - \alpha^*\hat a) \sigma_z/2\sqrt{2})$, with $\hat a$ the annihilation operator of the cavity mode, $\sigma_z$ the Pauli matrix, and $\alpha = (q+ip)$ the displacement in phase space. 
    
    An experimentally demonstrated approach to stabilize GKP states is the small-big-small (sBs) protocol \cite{Royer:2020,DeNeeve:2022,Sivak:2023}. 
    This protocol aims to mimic, via trotterization, a bath engineering process whose unique steady state is the qunaught state. The sBs protocol achieves this stabilization by a series of unitaries composed of control displacements intertwined by qubit reset onto the  eigenstate $\ket{+}$ of $\hat\sigma_x$, see \cref{fig:setup} c) and d). The unitaries applied between resets are $\hat U_{x,\Delta}$, with $x\in \{q, p\}$. We show this unitaries in circuit form in \cref{fig:setup} d), where $\varphi_q = 0$, $\varphi_p = \pi/2$, $s_\Delta = \sinh(\Delta^2)$, and $c_\Delta = \cosh(\Delta^2)$. The fact that $s_\Delta\ll c_\Delta$, gives the protocol its name, as a large control displacement is done in between two small ones. For an explanation on the derivation of these unitaries, and further details on the sBs protocol see \cref{Appendix:sBs}. Below, we proceed to describe how the protocol is used for the estimation of displacements, and the metrological performance it achieves. 

    \section{Results}
    \label{sec: Results}
    
    \subsection{Description of the protocol}
    \label{sec:description-protocol}
    
    The sBs metrology protocol is as follows. After having prepared the cavity in the finite-energy qunaught state $\rho_{\#_\Delta}=\ketbra{\#_\Delta}$, a displacement channel acts on it transforming the state onto $\rho(q_0,p_0) = \hat D(\beta)\rho_{\#_\Delta}\hat D(\beta)^\dagger$, with $\beta = (q_0+ip_0)/\sqrt{2}$, and $\hat D(\beta) = \exp(\beta \hat a^\dagger- \beta^*\hat a)$. The displacements of each quadrature $q_0$, $p_0$, are the variables we aim to estimate.  Then, the sBs stabilization protocol is repeatedly applied, where the reset of the qubit is substituted by measurement and feedback, see \cref{fig:setup} e). 
    With each qubit measurement, we gain one bit of information on the initial displacement of the oscillator state. Below, we refer to this bit acquisition stage of the protocol as the sBs bit acquisition.
    
    The unitaries $\hat U_{x,\Delta}$ with $x=q,p$ depicted in \cref{fig:setup}~(d) strongly resemble the textbook phase estimation circuit, except for the small displacements. These small displacements are the ones responsible for the stabilization of the cavity towards the finite-energy qunaught state $\rho_{\#_\Delta}$. In comparison to directly applying the textbook phase-estimation, where the number of photons would exponentially grow as more bits are acquired, adding these small displacements bounds the number of photons, and preserves the nongaussian structure of the grid in the presence of noise. This comes at the cost of a backaction kicking the oscillator back to the center of phase-space after each round, making subsequent bits carry less information about the initial displacements $q_0$, $p_0$.

    \begin{figure}[t]
        \centering
        \includegraphics[width=\linewidth]{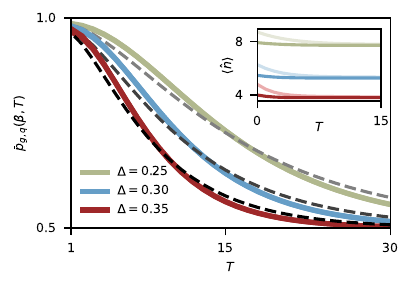}
        \caption{\textbf{Noiseless dynamics of the protocol.} Probabilities of measuring the qubit in its ground state $\bar p_{g,q}(\beta,T)$ calculated with the average recovery map, after initial displacement $\beta=(q_0+ip_0)/\sqrt{2}$ with $q_0=l/4$ and $p_0\in \{0,l/2\}$, for different envelope widths $\sim 1/\Delta$.  Dashed lines correspond to \cref{eq:p_xg}. Parameters $a_1,a_2$ are fitted, with $a_1=0.4$ the best fit at $T=0$ for all $\Delta$, and $a_2\simeq 1.24$ fitted for each $\Delta$. The inset shows the number of photons as a function of the number of rounds $T$. Lighter (darker) lines are used for a starting displacement of $p_0=l/4$ ($p_0=0$).}
        \label{fig:ideal-dynamics}
    \end{figure}
    
    Concretely, the positive operator valued measure (POVM) of the sBs protocol is constructed from the Kraus operators $\hat K_{x,g/e} = \bra{g/e}\hat U_{x,\Delta}'\ket{+}$, where the prime denotes that the last control displacement is not included in the unitary. For the explicit form of these Kraus operators, see \cref{Appendix:sBs}. After $T$ rounds, where one round composes the acquiring of two bits, one for each quadrature, the bitstring obtained from the qubit measurements is $\vb b =\bigl(b_1^q,b_1^p,\dots,b_T^q,b_T^p\bigr)$. The associated measurement channel to this bitstring is $\mathcal E_{\vb b}\rho = \hat K_{\vb{b}}\rho \hat K_{\vb{b}}^\dagger$, where $\hat K_{\vb{b}} = \prod_{j=1}^T e^{(-1)^{b_j^p} is_\Delta l\hat q/4} \hat K_{p,b_j^p}\hat e^{-(-1)^{b_j^q} is_\Delta l\hat p/4}\hat K_{q,b_j^q}$. In this expression, the displacements between the Kraus operators are the small displacement feedback dependent on the outcome of the qubit measurements.
    
    Due to the backaction kicking the oscillator back to the center of phase space, standard simplifications done in the analysis of quantum phase estimation algorithms are unavailable, and simple analytical expressions for the probability of a bit string, $p(\boldsymbol b|q_0,p_0) = \tr\bigl(\hat K_{\vb{b}}^\dagger \hat K_{\vb b}\rho(q_0,p_0)\bigr)$, are out of reach. However, great simplification is achieved when considering the averaged measurement probabilities of the qubit at round $T$. These probabilities are $\bar p_{x,g}(q_0,p_0,T) \equiv \tr\bigl[\hat K_{x,g}^\dagger\hat K_{x,g}\mathcal E_{T-1} \hat \rho(\beta)\bigr]$, with $\mathcal E_{T-1}$ the averaged measurement channel after $T-1$ rounds $\mathcal E_{T-1} = \sum_{\vb b} \mathcal E_{\vb b}$, and are approximated by
    \begin{equation}
    \label{eq:p_xg}
        \bar p_{g/e,x}(q_0,p_0,T) \simeq \frac{1}{2}\Bigl[1\pm e^{-a_1\Delta^2}\sin\bigl(lc_\Delta x_0e^{-a_2\Delta^2(T-1)}\bigr)\Bigr],
    \end{equation}
    with fitted parameters $a_1,a_2$. 
    We numerically obtain these averaged measurement probabilities using the fock basis, with cutoff Hilbert space dimension of 140. In \cref{fig:ideal-dynamics}, we plot these probabilities (full lines) for a displacement in the $q$ quadrature $q_0=l/4$, and displacements in the $p$ quadrature $p_0 \in \{0,l/4\}$, for different envelopes with $\Delta\in \{0.25,0.30,0.35\}$. We also plot the probability given by \cref{eq:p_xg} (dashed lines), with qualitative agreement. Note that only one full line per envelope width $\Delta$ is perceivable, as qubit measurement probabilities after application of $\hat U_{q,\Delta}'$ are nearly independent of the $p$ quadrature. In other words, the bits $b_t^q$ ($b_t^p$) carry information only about $q_0$ ($p_0$). Hence, the multivariate estimation problem, is reduced almost exactly to two independent ones. Additionally, in the inset we plot the mean photon number as a function of the number of rounds $T$. As shown there, the mean photon number remains bounded, and near the steady-state value at all times. For $p_0=l/2$ (lighter lines), initially more photons are in the oscillator in comparison with $p_0=0$ (darker lines).

    To summarize, we highlight three features of the sBs metrology protocol. First, the stabilization of the quadratures is nearly independent from one another. This implies that we can separate the multivariate estimation problem in two independent ones, with negligible loss in precision. Second, the decay rate is a function of the envelope width $\sim 1/\Delta$. The larger the envelope width, the slower the decay rate, and more informative are the acquired bits at later rounds. This gain in precision is consistent with the fact that the larger the envelope width is, the more photons are in the sensor state. Third, both the mean number of photons, and the phase space grid structure are preserved during the protocol. This hints at the possibility of recycling the resulting oscillator state for a new round of sensing, without the need of emptying the cavity and restarting the whole protocol. Below, we proceed to characterize the metrological performance. For more details in the description of the protocol, 
    including a quantitative explanation on the independence of the quadratures, and a more accurate expression for the averaged probabilities, 
    see \cref{Appendix:sBs}.


    \subsection{Metrological potential}
    \label{sec:metrological-potential}
    
    Here, we study the metrology potential of the sBs protocol in estimating the initial displacement $(q_0,p_0)$. From the analysis above, we can split the problem in two independent ones, one for each quadrature. This is due to bitstrings $\boldsymbol b_q$ ($\boldsymbol b_p$) containing information almost only on the $q$ ($p$) quadrature displacement $q_0$ ($p_0$). First, we define the measures of metrological performance. One of the main quantifiers available is the mean-square error, bounded for any estimator $\tilde q(\boldsymbol b)$ by the Cramer-Rao bound \cite{van2004detection},
    \begin{equation}
    \label{eq:CRB}
        \Ev[\boldsymbol(\tilde q(\boldsymbol b)-q_0\boldsymbol)^2]_{q_0}-B(q_0)^2\geq \frac{\{\partial_{q_0}\Ev[\tilde q(\boldsymbol b)]_{q_0}\}^2}{F[p(\boldsymbol b|q_0)]},
    \end{equation}
    where $\Ev[f(\boldsymbol b)]_{q_0}=\sum_{\boldsymbol b}p(\boldsymbol b|q_0)f(\boldsymbol b)$, $\Ev[\boldsymbol(\tilde q(\boldsymbol b)-q_0\boldsymbol)^2]_{q_0}\equiv \delta \tilde Q^2$ is the mean-squared error, $F\{p(\boldsymbol b|q_0)\} = \Ev\{[\partial_{q_0}\ln p(\boldsymbol b|q_0)]^2\}$ is the Fisher information, and $B(q_0)=(\Ev[\tilde q(\boldsymbol b)]_{q_0}-q_0)$ is the bias. Operationally, the average of the mean-squared error over the prior range quantifies the single-shot accuracy of the estimation. In complement, if the interest is not on the single-shot error, but on the sensing accuracy, 
    the figure of interest is the sensitivity, defined as $\Delta \tilde q_0\equiv \sqrt{\Ev[\boldsymbol(\tilde q(\boldsymbol b)-q_0\boldsymbol)^2]_{q_0}}/|\partial_{q_0}\Ev[\tilde q(\boldsymbol b)]_{q_0}|$. In other words, if the channel generating the displacement is stochastic or the access to it is limited, and the goal is to determine a rough value of the displacement in each shot, the average mean-squared error is an adequate figure of merit. On the other hand, if access to a deterministic displacement channel is unrestricted, and the objective is to distinguish the exact value of the displacement from its immediate neighbors, i.e. distinguish $q_0$ ($p_0$) from $q_0+\epsilon$ ($p_0+\epsilon$) for small $\epsilon$, the sensitivity is a more informative quantifier of metrological performance. 

    \begin{figure}[t]
        \centering
        \includegraphics[width=\linewidth]{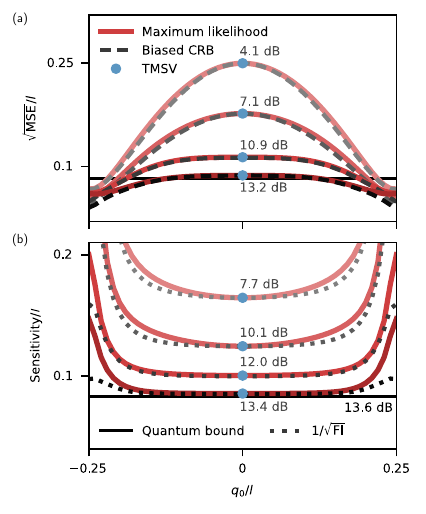}
        \caption{\textbf{Noiseless local metrological performance.} a) Root of the mean-square error, b) sensitivity of the protocol with maximum likelihood estimation as a function of the initial displacement $q_0 \in [-l/4,l/4]$. Red solid lines are the values obtained with maximum likelihood estimation. Lighter to darker lines represent 1, 2, 4 and 10 bits per quadrature respectively. Dark dashed lines correspond to the biased Cramer-Rao bound, RHS of \cref{eq:CRB}. Dotted lines are the classical bound on the sensitivity $1/\sqrt{F\{p(\boldsymbol b|q_0)\}}$. Solid black line is the quantum bound on sensitivity $1/\sqrt{4\bar n+2}$, see text for explanation. The legends next to the blue dots show the squeezing needed to achieve the highest (lowest) mean-square error (sensitivity) using two-mode squeezed vacuum states. The envelope width is set by $\Delta=0.3$.}
        \label{fig:metrological-potential}
    \end{figure}

    In what follows, we show the performance obtained by the protocol under both quantifiers, the mean-square error and the sensitivity. We start by focusing on the metrological performance under a flat prior $q_0,p_0 \in [-l/4,l/4]$. This choice of prior is standard if the displacements are believed to be small, and no further information is available. As estimation method, we use maximum likelihood estimation. In \cref{fig:metrological-potential}, we plot the square root of the mean-square error and sensitivity achieved after collecting 1, 2, 4 and 10 bits per quadrature (light to dark full red lines), as a function of the initial displacement $q_0 \in [-l/4,l/4]$. We also plot the Cramer-Rao bound for unbiased estimators (black dashed lines), and the quantum limit on the performance given by the multivariate quantum Cramer-Rao bound (full black lines). In addition, we mark with blue dots and accompanying legend, the squeezing level in units of dB necessary for two-mode squeezed vacuum states to achieve the same performance. For details on how we obtain the estimators, see \cref{appendix:noiseless}.     
    
    When a single bit is acquired, if the outcome of the qubit measurement is $g$ ($e$) the maximum likelihood estimator is $l/4$ ($-l/4$). This explains the observed small value of the mean-square error close to these values of $q_0$, and is limited by the contrast in front of the sine term in \cref{eq:p_xg}. On the other hand, near $q_0/l = 0$ the error is the half-width of the prior $l/4$. After acquiring 2 bits of information, the estimators are $\tilde q(gg/ee)=\pm l/4$ and $\tilde q(ge/eg)=\pm 0.025$, and thus the error near $q_0/l = 0$ is reduced. Acquiring more bits continues this process of filling the gaps between estimators of the previous round and, for the first few bits gathered, the accuracy of the estimation improves nearly linearly with the number of bits. However, as more bits are collected, the improvement in sensing accuracy per bit slows down, due to the tradeoff between stabilization and signal acquisition introduced by the small displacements of the sBs protocol. Notably, the mean-square error achieved surpasses the Cramer-Rao bound for unbiased estimators (solid dashed lines) at values of $q_0$ near $l/4$. This is possible, because here the maximum likelihood estimator turns out to be biased. Importantly, independently of the chosen estimator, and contrary to the mean-square error, the sensitivity is always bounded by $\Delta \tilde q_0\geq 1/\sqrt{F\{p(\boldsymbol b|q_0)\}}$. 

    Remarkably, a key observation from \cref{fig:metrological-potential} is that 
    as more bits are acquired, the sensitivity approaches the quantum bound $\Delta \tilde q_0\geq 1/\sqrt{4\bar n+2}$, given by the multivariate quantum Cramer-Rao bound  (full horizontal line). For example, for an envelope with $\Delta=0.3$, the number of photons in the cavity is $\bar n\simeq 2.6$. At this number of photons the quantum limit, when measured in units of dB of two-mode squeezed vacuum states, is of $13.6$ dB. When 10 bits per quadrature are acquired (20 measurements in total), the resulting sensitivity is equivalent to the one achieved with two-mode squeezed vacuum states with $13.4$ dB of squeezing, $0.2$ dB shy of the quantum limit. To achieve this sensitivity with two-mode squeezed vacuum states, $\bar n \simeq 5.0$ photons per quadrature would be needed. Nearly twice the efficiency per number of photons is reached. Notably, one bit per quadrature suffices to beat the classical sensitivity achieved with coherent states and heterodyne detection. Moreover, we find that the achieved sensitivity nearly saturates, but does not surpass, the upper bound on the Holevo bound given by $\sqrt{1+1/(8\bar n+4)}/\sqrt{4\bar n+2}$. For a detailed discussion on this last technical point, see \cref{appendix:quantum-bound} where we derive the upper bound, and \cref{appendix:noiseless} where we show the sensitivity achieved after acquiring $10$ bits per quadrature as a function of the envelope width, and compare it with both, the upper bound on the Holevo bound, and the multivariate quantum Cramer-Rao bound. In addition, for details on the sensitivity convergence with the number of acquired bits, see \cref{appendix:noiseless}. 
    
    \begin{figure}[t]
        \centering
        \includegraphics[width=.9\linewidth]{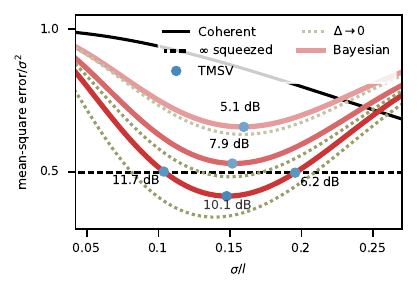}
        \caption{\textbf{Noiseless metrological performance with Gaussian priors.} Mean-squared error of Bayesian estimation as a function of the prior standard deviation. Only results for the $q$ quadrature are shown. Lighter to darker red lines are the values obtained with 1, 2 and 4 bits per quadrature. Dashed green lines are obtained analytically assuming infinite-energy with 1 and 2 bits per quadrature. Solid (dashed) black line is the coherent (Gaussian) limit of displacement estimation. The legends next to the blue dots show the squeezing needed to achieve the same mean-square error using two-mode squeezed vacuum states.}
        \label{fig:bayesian-noiseless}
    \end{figure}
    
    Going beyond the finite $[-l/4,l/4]$ range, we discuss the metrological performance when the prior is Gaussian $P[q_0,p_0] =\mathcal G_\sigma(q_0)\mathcal G_\sigma(p_0)$. This scenario mathematically arises either due to incomplete prior information or an inherent stochasticity of the signal, e.g.~when a force $F(t)= Af(t)$ linearly coupled to the oscillator fluctuates randomly in time with autocorrelation function $\delta(t-t')$. Hence, we take the averaged mean-square error as the figure of merit. We adopt a Bayesian inference approach, and use as estimator the conditional mean of the posterior. This choice of estimator is ideal when the objective is minimizing the averaged mean-square error \cite{van2004detection}. In \cref{fig:bayesian-noiseless}, we show the mean-square error of the $q$ quadrature normalized by the prior variance $\sigma^2$, as a function of the prior standard deviation $\sigma$, when 1, 2 and 4 bits per quadrature are collected, for an envelope with $\Delta=0.30$. In the limit of $\Delta\to 0$, we analytically find that the Gaussian limit can be surpassed with as few as 2 bits per quadrature. However, with the range of envelopes we numerically explore $\Delta\in [0.25,0.4]$, we find that in practice 3 to 4 bits are required. Acquiring more bits increases the prior range where the Gaussian limit is surpassed, up to a convergent value depending on the envelope. 
    Finally, we note that the smaller the prior variance, the better the sBs metrology protocol fares in comparison with two-mode squeezed vacuum states.

    \subsection{Backaction evading sensor}
    \label{sec:backaction}

    In the presence of an itinerant signal, possibly of varying amplitude and phase, the sensor must be repeatedly probed for signatures of the signal. Backaction evading measurements are ideal for such sensors. These measurements, as defined by Caves in Ref.~\cite{caves2020reframing}, are those where the measurement itself prepares the sensor in an appropriate quantum state, ready for the next probing interval and the next measurement. The sBs metrology protocol is a backaction evading sensor. This is so, because as bits are acquired, the measurement backaction drives the cavity state back to the qunaught GKP state. As such, this sensor is well suited for detecting and estimating \emph{itinerant signals}, for example fluctuating forces.

    \begin{figure}[t]
        \centering
        \includegraphics[width=\linewidth]{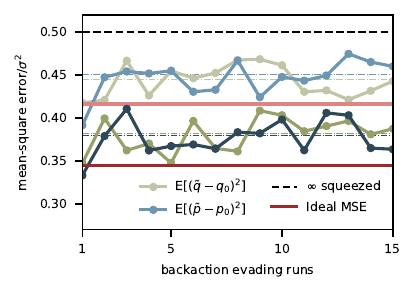}
        \caption{\textbf{Noiseless backaction evading performance.} Mean-squared error of Bayesian estimation with continuous operation of the sensor and total bit budget of 12 bits per quadrature. See text for a detailed description of the protocol. Lighter to darker green (blue) lines are the mean-square errors obtained with 4000 repetitions of the protocol for the $q$ displacement estimation ($p$) with 4, and 8 bits per quadrature acquired. Red light (dark) solid line is the mean-square error obtained when the sensor state is the finite-energy qunaught state and 4 (8) bits per quadrature are acquired. Dashed black line is the Gaussian limit of displacement estimation. Dotted-dashed lines, added as a guide to the eyes, correspond to the average of the mean-square errors from the second run forward.}
        \label{fig:backaction}
    \end{figure}

    To test this backaction evading property we consider the following protocol. After the initial displacement, a bitstring $\boldsymbol{b}$ with $T$ bits per quadrature is obtained to estimate $\tilde q_0(\boldsymbol b^q)$ and $\tilde p_0(\boldsymbol b^p)$. Next, we apply a displacement $\hat D(-\beta_r)$ with $\beta_r=[\tilde q_0(\boldsymbol b_q)+i\tilde p_0 (\boldsymbol b_p)]/\sqrt{2}$. To correct for estimation errors, we follow with $M$ rounds of autonomous sBs, where the qubit is reset instead of measured. We refer to this sequence as the backaction evading protocol.
    
    Numerically, we test this by starting the protocol in the sensor state $\rho_{\#}=\ketbra{\#}$, and applying a random displacement sampled from a Gaussian prior $P[q_0,p_0]= \mathcal{G}_\sigma(q_0)\mathcal{G}_\sigma(p_0)$. Then, the backaction evading protocol is performed, with resulting bitstring $\boldsymbol{b}_1$, Bayesian estimators $\tilde q_0(\boldsymbol b_1^q)$, $\tilde p_0(\boldsymbol b_1^p)$, and state $\rho_1 = \mathcal{R}_{\boldsymbol b_1} \rho_{\#}$. This output state $\rho_1$ is again randomly displaced and used as input for the next iteration of the backaction evading protocol. We repeat this sequence $N$ times. The parameters $T$ and $M$ are chosen to maximize the fidelity of the final state with the qunaught finite-energy state. This motivates a choice of $M$ that is as large as possible; however, these are uninformative bits, and should be kept at a minimum. For example, choosing an envelope with $\Delta=0.3$, standard deviation of the prior $\sigma = 0.15l$, and a total bit budget of $T+M=12$ we find that setting $T=8$ achieves the largest average recovery fidelity of the final state with the initial one when weighted by the Gaussian prior $p(q_0)=\mathcal G_\sigma(q_0)$. Here, the choice of bit budget $T+M=12$ is arbitrary, beyond it being large enough to allow the protocol surpassing the Gaussian limit for several different pairs $\{T,M\}$. In an applied setting, the target accuracy, and the sensitivity per unit of time are figures of merit that can be used to optimize $T$ and $M$. This optimization will depend on the experimental platform and particularities of the experiment, and we do not explore it further here. For details on the achieved average fidelity, see \cref{appendix:noiseless}. 
    
    \Cref{fig:backaction} shows the mean-square error in estimating the $q$ quadrature and $p$ quadrature displacements (full green and blue lines) with $N=15$, $T=8$ and $M\in \{4,8\}$ (darker lines for $M=8$), averaged over 4000 samples of the whole sequence.
    As benchmark, the full red line shows the performance achieved starting from the finite-energy qunaught state, in the asymptotic limit of samples. After only two backaction evading runs, the protocol reaches a steady mean-square error (dashed lines show the average performance achieved for $N\geq 2$) and outperforms the single-mode Gaussian limit, confirming that the sBs metrology protocol effectively functions as a backaction-evading sensor. The decrease in performance in comparison with the ideal case reflects, on average, a deviation of the sensor state from the center of phase-space at the end of each run due to estimation errors and the slow recovery rate of the sBs protocol. In \cref{sec:discussion}, we point out possible strategies to decrease this deviation and improve the information rate per unit time of the protocol.
    
    \subsection{Noise analysis}
    \label{sec:noise}

    We now consider the effect of decoherence on the performance of the metrology protocol. Most of the errors in sBs stabilization occur in the big conditional displacements, where a qubit relaxation event propagates to the cavity and can cause a large displacement on the sensor state \cite{Royer:2020}. Hence, we use a simplified noise model where all qubit rotations and small conditional displacements are perfect and free of noise. This assumption is appropriate since those operations can be made much faster than decoherence timescales of the system. To account for both qubit and oscillator relaxation and dephasing noise channels during the big conditional displacements, we numerically solve the Lindblad master equation $\dot{\rho} = -i [\hat{H}_{CD}(\beta,T_{CD}), \hat{\rho}] + \sum_i \mathcal{D}(\hat{c}_i)\hat{\rho}$, where $\mathcal D(\hat c_i)\hat \rho = \hat c_i\hat \rho \hat c_i^\dagger -\{\hat \rho, \hat c_i^\dagger \hat c_i\}/2$, with 
    \begin{equation}
        \hat{H}_{CD} (\beta,T_{CD}) = \frac{1}{T_{CD}}\frac{1}{2\sqrt{2}} (i \beta \hat{a}^\dagger - i\beta^* \hat{a})\otimes \hat{\sigma}^z.
    \end{equation}
    We use the collapse operators, $\hat{c}_1 = \hat{\sigma}^- / \sqrt{T_{1,q}}$, $\hat{c}_2= (1-\hat{\sigma}^z) / \sqrt{2T_{\phi,q}}$, $\hat{c}_3 = \hat{a} / \sqrt{T_{1,c}}$, and $\hat{c}_4 = \sqrt{2}\hat{a}^\dagger \hat{a} /\sqrt{T_{\phi,c}}$, where $T_{1,q(c)}$ is the relaxation time and $T_{\phi,q(c)}$ the pure dephasing time of the qubit (cavity). The pure dephasing time is given by $T_{\phi,q(c)} = (1/T_{2,q(c)} - 1/2T_{1,q(c)})^{-1}$, where $T_{2,q(c)}$ is the coherence time of the qubit (cavity). 
    
    To evaluate the metrological performance of the protocol, we use as reference the experimentally measured lifetimes of \textcite{Sivak:2023}, $T_0 = \{T_{1,q}^0, T_{2,q}^0, T_{1,c}^0, T_{2,c}^0 \}$ with values $T_{1,q}^0 = 280 \ \mu$s, $T_{2,q}^0 = 240 \ \mu$s, $T_{1,c}^0 = 610 \ \mu$s, $T_{2,c}^0 = 980 \ \mu$s, and a gate time of the big conditional displacement of $T_{CD} = 500$ ns.  We compare the metrological performance for different noise levels set by a multiplication factor $\eta$ for all coherence times in $T_0$. In other words, we use the lifetimes $T_\eta =\{\eta T_{1,q}^0,\eta T_{2,q}^0,\eta T_{1,c}^0, \eta T_{2,c}^0\}$. 

    \begin{figure}[t]
        \centering
        \includegraphics[width=\linewidth]{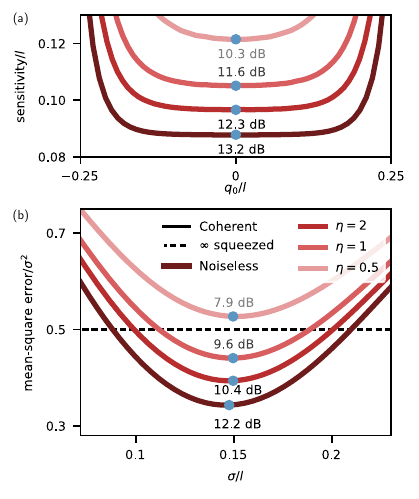}
        \caption{\textbf{Metrological performance in presence of noise.} a) Sensitivity obtained with maximum likelihood estimation and flat prior $q_0 \in [ -l/4, +l/4]$, b) mean-square error obtained with Bayesian estimation, when 8 bits per quadrature are gathered. Only results for the $q$ quadrature are shown. Lighter to darker red solid lines correspond to the sensitivity and mean-square error obtained with decreasing degree of noise. Blue dots and legends show the required squeezing to achieve the same sensitivity (mean-square error) using noiseless two-mode squeezed vacuum states. In the bottom, black solid (dashed) line is the coherent (Gaussian) limit of displacement estimation.}
        \label{fig:noise-performance}
    \end{figure}
    
    In \cref{fig:noise-performance}, we show the sensitivity obtained assuming a flat prior $q_0,p_0 \in [-l/4,l/4]$ while using maximum likelihood estimation, and the mean-square error as a function of the Gaussian prior standard deviation while using Bayesian estimation, for different noise levels. We obtain these results by using as initial state the steady state of the sBs protocol in presence of noise $\rho_{\#_\Delta,\eta}$, where $\eta$ quantifies the strength of the noise used. Then, the sBs bit acquisition is performed in the presence of noise. We find that, as in the noiseless case, both quadratures are nearly independent, and we show only results for the $q$ quadrature when 8 bits per quadrature are gathered setting $p_0=0$. The sensitivity at the center ($q_0=0$) is less affected by noise, with a sharp decrease in performance near the borders of the prior. For both sensitivity with maximum-likelihood estimation, and mean-square error with Bayesian estimation, the difference in performance from a multiplication factor on $T_0$ of $2$ to 1, and from 1 to 0.5 shows that as the noise increases, the performance degradation gradient becomes steeper. This is mostly due to a sharp decrease in the fidelity between the noisy sBs steady state and the finite-energy qunaught state. Thus, improved state preparation \cite{Eickbusch:2022,singh2025towards}, can greatly improve the achieved performance.
    
    Our results show that while performance is degraded in the presence of noise, and approaching the quantum limit is no longer possible with the sBs protocol, sensitivities approaching 10 dB, and unconditionally surpassing the Gaussian limit, is within reach with state-of-the-art hardware. For details in the metrological performance in the estimation of the $p$ quadrature, the state preparation fidelity with respect the ideal sensor state, a breakdown of the effects of each type of noise, and the backaction performance in presence of noise see \cref{appendix-noise}.

    \section{Discussion and Summary}
    \label{sec:discussion}

    We showed how a quantum error correction inspired stabilizing dynamics on a bosonic mode, can be used for the metrology of displacements in both quadratures with a sensitivity approaching the quantum limit given by the multivariate quantum Cramer-Rao bound \cite{braunstein1994statistical,genoni2013optimal,Duivenvoorden:2017,pezze2025advances}. Concretely, using an auxiliary qubit for control and measurement, a GKP state is stabilized while extracting information on the displacement, thus being a backaction evading sensor as defined by Caves in Ref.~\cite{caves2020reframing}. Even in the presence of noise, we showed how any single mode Gaussian strategy for displacement sensing can be unconditionally beaten, even if the latter is allowed an infinite number of photons \cite{Hanamura:2021}. Moreover, our results hint that sensitivities reaching and possibly surpassing 10 dB can be obtained in current state-of-the-art microwave hardware, matching the performance obtained in recent experiments at the optical regime \cite{zander2021full}.
    Our work also provides a valuable case study in quantum multivariate estimation. Specifically, it shows how is possible to approach the multivariate quantum Cramer-Rao bound, even if the weak-commutativity condition is not satisfied \cite{matsumoto2002new, pezze2025advances}. This is the case due to the small ratio between the Uhlmann curvature and the quantum Fisher information matrix. Concretely, this ratio puts an upper limit on the Holevo bound $ \mathcal B_H\leq  1/(2n+1)+1/4(2n+1)^2$ \cite{carollo2019quantumness}, which then approaches the QCRB as $1/16n^2$, where $n$ is the mean number of photons in the sensor state. For details on the derivation, see \cref{appendix:quantum-bound}. 
    Additionally, our work shows that the sBs protocol for stabilizing GKP states, is nearly as slow as possible, as if it were slower in correcting displacement errors than the quantum bound could be surpassed. As such, our work provides an example of how quantum metrology lower bounds stabilization speeds. Moreover, although the experimental parameters we used in our noise analysis are drawn from superconducting microwave experiments, the protocol is platform independent, and can in principle be performed in any platform realizing the qubit-oscillator pair, such as trapped ions or mechanical oscillators~\cite{arrangoiz2019resolving}. 

    Moving forward, many open questions remain. A pathway to improve the information rate per unit time is alternating between the sBs protocol for bit recollection, and another faster stabilizing method to finish bringing the state back at the center of phase-space \cite{sellem2025dissipative}. Going beyond the estimation of the displacement in each shot, the learning of the channel inducing the displacement is a natural continuation. In Ref.~\cite{Oh:2024} the authors showed how two-mode squeezed vacuum states can be used to gain an exponential advantage with respect to any entanglement-free strategy in learning a random displacement channel acting on $N$ oscillators. In analogy with the estimation case, if prior information assures that the displacement amplitudes are well below the vacuum width, then we conjecture multimode GKP states \cite{gottesman2001encoding,harrington2001achievable,Royer:2022,conrad2022gottesman} will not only provide an exponential advantage over the entanglement-free strategies, but a constant factor improvement in the number of photons needed in comparison to two-mode squeezed vacuum states, an improvement that can prove to be very significant in practice. 
    
    Additionally, here we have not tackled amplitude estimation. In particular, in quantum waveform estimation, if the waveform is weak and stochastic, the sensing problem becomes the detection of a small Gaussian noise \cite{shi2023ultimate,gardner2024stochastic}. In such case, the relevant quantity is the amplitude of the displacement, or, in other words, the excess energy coming from the stochastic source, while the phase can be ignored \cite{gorecki2022quantum,grochowski2025optimal}.
    Concretely, in Ref.~\cite{gardner2024stochastic} the authors showed that in the presence of noise and in the limit of small signal, GKP states outperform two-mode squeezed vacuum states in this task, with similar performance only if the memory mode is pure (noiseless). Additionally, in the relevant and closely related problem of dark-matter detection, \cite{shi2023ultimate} it was shown that two-mode squeezed vacuum states are optimal sensor states when the memory mode is noiseless, up to a factor of two if the photons in the memory mode are counted. Once again, in analogy with the displacement problem, in presence of noise one may expect similar or better performance from GKP states. Thus, a careful analysis of the potential stabilized GKP states have for weak-signal detection scenario seems desirable, as the backaction evading properties of the sensor make it ideal for such task. However, amplitude estimation is a single-parameter estimation problem with challenges differing from that of measuring displacements. Concretely, the interest in dark-matter searches and weak stochastic waveform detection is set on detecting very small signals, much smaller than the vacuum width \cite{bertone2018new,Backes:2021,dixit2021searching,shi2023ultimate,gardner2024stochastic}. As such, a comprehensive analysis of the potential stabilized GKP states have for weak-signal detection remains open.
     
        

        

    \section*{Acknowledgements}
    
    The authors are grateful to Cristóbal Lledó and Manuel H. Muñoz-Arias for useful discussions, and Simon Richer for a careful reading of the manuscript. This material is based upon work supported by the U.S. Department of Energy, Office of Science, National Quantum Information Science Research Centers, Quantum Systems Accelerator. Additional support is acknowledge from NSERC, the Ministère de l’Économie et de l’Innovation du Québec, the Fonds de recherche du Québec – Nature et technologie and the Canada First Research Excellence Fund.

    \appendix    

    \section{Background review}
    \label{appendix:background}

    Here, we discuss relevant background on displacement sensing, and quantum-error correction assisted metrology. In each subsection, we highlight the most notable features of the sBs metrology protocol, in connection with the previous literature.
    
    \subsection{Displacement sensing.} 
    \label{appendix:displacement-sensing}
    
    When only Gaussian states and resources are allowed, without prior information the best single-mode strategy in the estimation of two-quadrature displacements is to prepare the oscillator in a coherent state followed by heterodyne measurement. In that case, the mean-square error in the estimation achieved is $\delta \tilde Q^2+\delta \tilde P^2 = 2$,  where $\delta \tilde Q^2$ ($\delta \tilde P^2$) is the mean-square error on the estimation of the $q$ ($p$) quadrature (here and henceforth $\hbar =1$) \cite{genoni2013optimal}. 
    When all quantum states and operations are allowed, the estimation precision is bounded by the Holevo bound \cite{holevo2011probabilistic}, which can always be saturated in the asymptotic limit. A looser bound is given by the by multivariate quantum Cramer-Rao bound (QCRB), which is given by the trace of the quantum Fisher information matrix $Q_M(\rho)$ \cite{helstrom1969quantum, pezze2025advances}. When only a single-mode is used as sensor, the resulting QCRB is $\delta \tilde Q^2+\delta \tilde P^2\geq 1/(2\bar n+1)$, where $\bar n = \mathrm{Tr}[ \hat{n}\rho]$ is the mean number of photons in the sensor state \cite{Duivenvoorden:2017}. This bound is saturated when $\bar n\to \infty$, where the gap between the Holevo bound and the QCRB vanishes \cite{matsumoto2002new, pezze2025advances, carollo2019quantumness}. For details on the derivation of QCRB, the conditions under which it can be saturated, and a detailed discussion on the gap between the Holevo and quantum Cramer-Rao bounds see \cref{appendix:quantum-bound} and Ref.~\cite{frigerio2025joint}. 
    
    Going beyond single-mode strategies, the optimal Gaussian strategy uses two-mode squeezed vacuum states, and has a mean-square error of $\delta \tilde Q^2+\delta \tilde P^2 = 2e^{-2r}$, where $r$ is the squeezing parameter \cite{genoni2013optimal,steinlechner2013quantum,bradshaw2018ultimate,caves2020reframing,zander2021full}. Exploiting the concept of quantum mechanics free subspaces, both quadratures can be estimated with unbounded precision, bypassing the uncertainty principle~\cite{tsang2012evading}. 
    Both the single-mode coherent state strategy and the use of two-mode squeezed vacuum have the advantage of having an estimation precision agnostic to the value of the displacement. Hence, in practice, when there is no prior information about the displacement these are the best strategies.
    
    
    When Gaussian priors are available, with probability distribution $P_0[Q,P] = \mathcal G_{\sigma}[Q]\mathcal G_{\sigma}[P]$, with $\mathcal G_{\sigma}[x] = e^{-x^2/2\sigma^2}/\sqrt{2\pi\sigma^2}$, the single-mode Gaussian limits  on the mean-square error $\delta \tilde Q^2+\delta \tilde P^2$ are \cite{Hanamura:2021},
    \begin{equation}
    \label{eq:gaussian}
        \begin{cases}
        \sigma^2, &\sigma<1\\
        \frac{2\sigma^2}{\sigma^2+1}, &\sigma\geq 1.
        \end{cases}
    \end{equation}
    The first case corresponds to subvacuum sensing, and the bound is saturated by infinitely squeezed states with homodyne detection. The second case, where the prior variance is greater than the vacuum noise, corresponds to classical sensing, and the bound is saturated by coherent states and heterodyne detection. In the same scenario, the optimal mean-square error on the estimation with two-mode squeezed vacuum is given by \cite{genoni2013optimal},
    \begin{equation}
    \label{eq:SU}
        \delta \tilde Q^2 + \delta \tilde P^2 = \frac{2e^{-2r}\sigma^2}{e^{-2r}+\sigma^2},
    \end{equation}
    where $r$ is the squeezing parameter in $\hat S(r)=e^{r(\hat a^2-\hat a^{\dagger 2})/2}$.
    The Gaussian limit for subvacuum sensing is beaten using two-mode squeezed vacuum states  
    when the squeezed quadratures variance is below the prior variance, \emph{i.e.} when $e^{-2r} < \sigma^2$. Additionally, we note that Ref.~\cite{Hanamura:2021} theoretically showed that the Gaussian limit can be beaten in an appropriate range of priors with postselection using a single mode fock states as sensor, in conjunction with a second mode also prepared in a fock state. Concretely, the postselection criteria there is keeping only displacements smaller than certain value. In the main text, we compared the performance of the sBs metrology protocol with the optimal precision achieved with two-mode squeezed vacuum states \cref{eq:SU}, and showed that it beats both the classical and subvacuum Gaussian limits in an appropriate range of priors without postselection.
    
    Beyond Gaussian priors, when the displacement is believed to be small, the use of flat priors confining the possible displacements to a finite range $q_0,p_0\in [-a,a]$ is common in practice. When $a=\sqrt{\pi/2}$, Ref.~\cite{Duivenvoorden:2017} showed how single-mode modular measurements \cite{terhal2016encoding} can be used to construct an estimator with mean-squared error $\delta \tilde Q^2+\delta \tilde P^2 = O(1/\sqrt{\bar n})$. In that case, measuring modular functions of the quadratures allows bypassing the uncertainty principle \cite{aharonov1969modular}. In the sBs metrology protocol, while ignoring noise, we numerically find that the sensitivity approaches the multivariate quantum Cramer-Rao bound $\delta \tilde q^2+\delta \tilde p^2 \geq 1/(2\bar n+1)$. This comes at the cost of restricting the prior to the range $q_0,p_0\in [-a,a]$ with $a=\sqrt{\pi/8}$. For details of the numerical convergence to the bound, see \cref{appendix:noiseless}. For two-modes, if only the photons of one of the modes are considered, two-mode squeezed vacuum states formally saturate the bound in the limit of infinite squeezing. However, if photons of both the sensor and the memory modes are counted, the sensitivity falls short of the quantum bound in photons per quadrature by a factor of 2.

    \subsection{QEC assisted metrology}
    \label{appendix:QEC-assisted-metrology}
    
    In QEC assisted metrology protocols, a fundamental tradeoff exists between sensitivity and robustness. In its simplest form, this is captured by the expression
    \begin{equation}
    \label{eq:QEC-metrology}
        (\mathcal R \circ \mathcal E_\Gamma \circ \Lambda_X)\rho = (\mathcal E_{g(\Gamma)} \circ \Lambda_{f(X)})\rho,
    \end{equation}
    where $\Lambda_X$ is the encoding quantum channel dependent on the parameter $X$ to be sensed, $\mathcal E_\Gamma$ is the noise channel dependent on the strength of the noise~$\Gamma$, and $\mathcal R$ is the recovery channel, with $\Lambda_0=\mathcal E_0 = I$. The ideal case is $f(X) = X$,~$g(\Gamma) =0$, that is, the encoding of the signal remains unchanged while all the noise is suppressed. If the encoding quantum channel is the unitary $\Lambda_X \dot = e^{-iX\hat G} \dot e^{iX\hat G} $, $P$ is the projector onto the code space, and~$\mathcal E = \sum_k \hat E_k\rho \hat E_k^\dagger$, the necessary and sufficient requirements to achieve the ideal QEC condition where all the noise is suppressed without affecting the signal, are $(i)$~$[\hat G, P] = 0$, and $(ii)$~$P\hat E_i\hat E_jP = A_{i,j}P$ with $A=(A_{i,j})$ a hermitian matrix \cite{kessler2014quantum}. For displacement sensing, this requires a code with infinite energy codewords. Hence, we are restricted to the situation where $g(\Gamma)=0$~and~$f(X)=X$ cannot be both satisfied simultaneously. Even so, it might be the case that $g(\Gamma)$,~$f(X)$ are such that we can still reach the Heisenberg limit, or, in this context, saturate the quantum Cramer-Rao bound.

    A necessary and sufficient condition for reaching the HL in the number of probes or measurement time is the `Hamiltonian not in Lindblad span’ (HNLS) \cite{zhou2018achieving}. Concretely, consider a Lindblad equation $\dot \rho = -iX[\hat G, \rho] + D[\hat L]\rho$, then a necessary and sufficient condition for reaching the HL, is that the generator $\hat H$ is not in the hermitian span of $\{I,\hat L, \hat L^\dagger, \hat L\hat L^\dagger\}\equiv S_H[\hat L]$. This condition captures the following intuition. If the signal and the noise do not have orthogonal components, then any recovery operation correcting the noise also erases the signal. However useful, this theorem as presented in Refs. \cite{zhou2018achieving,zhou2021error}, does not apply to the sBs metrology protocol. This is so because, in the standard approach to QEC metrology, the sensor experiences an effective evolution within the codespace. On the contrary, in the protocol we present there is no such codespace, as we are using the qunaught GKP state. Nonetheless, useful intuition can be gained by it as we now discuss. 
    
    In the sBs metrology protocol, the qunaught GKP state is prepared by trotterizing a Lindblad dynamics $\dot \rho/\Gamma = D[\hat d_{q,\Delta}]\rho +D[\hat d_{p,\Delta}]\rho$ where $\hat d_{x,\Delta} = \ln \hat T_{x,\Delta}$, and $\Gamma$ is the decay rate. For infinite energy $\Delta = 0$, the generators of displacement, the linear span of $\{\hat q, \hat p\}$, are not in the Lindblad span of the dissipators, and hence, the displacement can be distinguished with infinite precision as it never decays. However, for any nonzero $\Delta$ the generators of displacements are in the hermitian span of these dissipators, and hence the stabilization progressively erases the encoded displacement while stabilizing the state back to the center of phase-space. The erasure rate depends on the number of photons used in the sensor state, and vanishes in the limit of an infinite number of photons. For finite-energy, we numerically find that is well fitted by an exponential decay. For details on this decay, see \cref{Appendix:sBs}. 
    
    \section{Quantum bound on two-quadrature displacement estimation}
    \label{appendix:quantum-bound}

    In this appendix, we provide a pedagogical explanation of the two-parameter quantum estimation problem \cite{Duivenvoorden:2017,pezze2025advances}. Concretely, we derive the multivariate quantum Cramer-Rao bound (QCRB) for two-quadrature displacement sensing. Moreover, we will show how the weak-commutativity condition and the measurement incompatibility quantified by the ratio between the Uhlmann curvature and the quantum Fisher information \cite{carollo2019quantumness,frigerio2025joint}, explains why the protocol we presented in the main approaches but does not saturate the QCRB. In doing so,  we formalize the intuition presented in Appendix A of Ref.~\cite{Duivenvoorden:2017}. We stress that throughout we focus on simplicity. For a rigorous mathematical treatment of multivariate quantum estimation, we refer the reader to Refs. \cite{holevo2011probabilistic,matsumoto2002new,carollo2019quantumness}, and to Ref.~\cite{pezze2025advances} for a broader overview. 

    In the two-parameter estimation problem, the estimation uncertainty is quantified by the covariance matrix of estimators, 
    \begin{equation}
    \label{appendix:eq-covariance-matrix}
        \mathsf{C}(\rho_{\boldsymbol x}, \mathcal{E}, \tilde{\boldsymbol x}) = \sum_{\vb b} p(\boldsymbol b|\boldsymbol x)[\boldsymbol x - \tilde{\boldsymbol x}(\boldsymbol b)][\boldsymbol x - \tilde{\boldsymbol x}(\vb b)]^T,
    \end{equation}
    where $\rho_{\boldsymbol x}$ is the family of density matrices parametrized by the displacement vector $\boldsymbol x = \mqty(q_0 & p_0)^T$, $\mathcal E$ is the measurement channel with POVMs $\{E_{\boldsymbol b}$\} and outcomes $\boldsymbol b$ (e.g. bitstrings), $\tilde{\boldsymbol x}(\boldsymbol b) = \mqty(\tilde q(\boldsymbol b) & \tilde p (\boldsymbol b))^T$ is the estimator vector, and $p(\boldsymbol b|\boldsymbol x)= \tr(E_{\boldsymbol b}\rho_{\boldsymbol x})$ is the probability of measuring $\boldsymbol b$ given a displacement $\boldsymbol x$. For unbiased estimators, the norm of this covariance matrix is lower bounded by the Fisher information matrix $\mathsf{F}(\rho_{\boldsymbol x},\mathcal E)$, which in turn is bounded by the quantum Fisher information matrix $\mathsf{F}_Q(\rho_{\boldsymbol x})$,
    \begin{equation}
    \label{appendix:mQCRB}
        \mathsf C (\rho_{\boldsymbol x}, \mathcal{E}, \tilde{\boldsymbol x})\geq \frac{\mathsf{F}^{-1}(\rho_{\boldsymbol x},\mathcal E)}{N}\geq \frac{\mathsf F_Q^{-1} (\rho_{\boldsymbol x})}{N},
    \end{equation}
    where $N$ is the number of repetitions of the measurement protocol. Operationally, this means that for any vector $\boldsymbol v$, it holds that $\boldsymbol v^T \mathsf C \boldsymbol v \geq \boldsymbol v^T \mathsf F^{-1} \boldsymbol v/N \geq \boldsymbol v^T \mathsf F_Q^{-1} \boldsymbol v/N$. For any $\boldsymbol v$, is always true that an optimal POVM saturating the rightmost inequality can be found in the asymptotic limit $N\to \infty$\cite{suzuki2020quantum}. 

    Generally, the cost function of the estimation is given by $\tr(\mathsf W \mathsf C)$, with the weight matrix $\mathsf W = \sum_j w_j \boldsymbol v_j\boldsymbol v_j^T$. For this general cost, we have the chain of inequalities,
    \begin{equation}
    \label{Appendix:holevo-inequality}
        \tr(\mathsf W \mathsf C)\geq \frac{\tr (\mathsf W \mathsf F^{-1})}{N} \geq \frac{\mathcal B_H(\rho_{\boldsymbol x}, \mathsf W)}{N}\geq \frac{\tr(\mathsf W \mathsf F_Q^{-1})}{N},
    \end{equation}
    where $\mathcal B_H(\rho_{\boldsymbol x})$ is the Holevo bound, always saturable in the asymptotic limit \cite{holevo2011probabilistic}. Below, we will proceed to upper bound the Holevo bound, and as such upper bound the achievable performance in the asymptotic limit \cite{pezze2025advances}. Following Ref.~\cite{carollo2019quantumness}, we do so by computing the QCRB, and then bounding its ratio with the Holevo bound.
    
    In the two-quadrature displacement estimation problem we are interested in the case where the cost is the sum of the mean-square errors  $\tr(\mathsf C) =\text{E}[(\tilde q(\boldsymbol b)-q_0)^2]+\text{E}[(\tilde p(\boldsymbol b)-p_0)^2]$, and $\mathsf W = \boldsymbol v_q\boldsymbol v_q^T+\boldsymbol v_p\boldsymbol v_p^T=I$ is the identity matrix. In virtue of the QCRB, rightmost inequality of \cref{Appendix:holevo-inequality}, this cost is bounded by $\tr(\mathsf F_Q^{-1})/N$. The quantum Fisher information matrix is defined as \cite{helstrom1969quantum,liu2020quantum},
    \begin{equation}
        \mathsf (\mathsf F_Q)_{\mu\nu} = \frac{1}{2}\tr(\rho_{\boldsymbol x} \{\hat L_{\mu},\hat L_{\nu}\}),
    \end{equation}
    where $\nu,\mu \in \{q,p\}$ and $\hat L_\mu$ are the symmetric logarithmic derivative (SLD) operators, implicitly defined as solutions of the equation $\partial_\mu \rho_{\boldsymbol x} = (\hat L_{\mu}\rho_{\boldsymbol x}+\rho_{\boldsymbol x}\hat L_{\mu})/2$. For pure states $\rho_{\boldsymbol x} = \ketbra{\psi_{\boldsymbol x}}$, and unitary encoding $\ket{\psi_{\boldsymbol x}} = \hat U(\boldsymbol x)\ket{\psi_0}$, with $\hat U(\boldsymbol x) = \exp(-iq_0\hat p+ip_0\hat q)$, the SLD operators evaluated at $q_0=p_0=0$ (locality of the QCRB allows this simplification) are $\hat L_q = 2i[-\hat p,\ketbra{\psi_0}]$, $\hat L_p = 2i[\hat q,\ketbra{\psi_0}]$. Then, the quantum Fisher information matrix is given by, 
    \begin{equation}
        \mathsf F_Q = 4\mqty(\langle\hat q^2\rangle & -\text{Cov}(\hat q, \hat p)\\ -\text{Cov}(\hat q, \hat p) & \langle\hat p^2\rangle),
    \end{equation}
    where $\ev{\bullet} = \mel{\psi_0}{\bullet}{\psi_0}$, and $\text{Cov}(\hat A, \hat B) = \langle\hat A\hat B+\hat B\hat A\rangle/2-\langle\hat A\rangle\langle\hat B\rangle$. The trace of the inverse is minimized when the offdiagonal terms vanish, hence, states with no covariance between the quadratures are the ones of interest. Example of such states are fock states, square GKP states and compass states \cite{zurek2001sub,Duivenvoorden:2017,shukla2023superposing}. Then, is direct to proof that the trace of the inverse is bounded by,
    \begin{equation}
        \tr\left(\mathsf F_Q^{-1}\right)\geq \frac{1}{2\bar n+1},
    \end{equation}
    with $\bar n = \ev{\hat n}$, the mean number of photons in the sensor state $\ket{\psi_0}$. 
    
    Having computed the QCRB bound, we now determine whether the Holevo bound saturates it. First, without explicit calculations, is straightforward to see that in the limit of infinite-energy the Holevo bound saturates the QCRB. In this limit, infinitely squeezed states in conjunction with homodyne measurements saturate the single quadrature QCRB. Instead of an homodyne measurement, a modular quadrature measurement would also saturate the bound, as the QCRB is a local quantity. Hence, as locally infinite-energy GKP states are infinitely squeezed states in both quadratures, measurements of the modular quadratures $\hat q_{[l]}$, $\hat p_{[l]}$ with $l=\sqrt{2\pi}$, saturate the bound. Beyond this idealized case, restricting ourselves to finite-energy, a necessary and sufficient condition for the Holevo bound to saturate the QCRB is \cite{matsumoto2002new,ragy2016compatibility}, 
    \begin{equation}
        \mathcal U_{\mu \nu} = \frac{1}{2i} \tr(\rho_{\boldsymbol x}[\hat L_{\mu},\hat L_{\nu}]) = 0,
    \end{equation}
    for all $\mu,\nu \in \{q,p\}$. The matrix $\mathcal{U}$ is refered to as the mean Ulhman curvature. This vanishing of the mean Ulhman curvature is known in the literature as \emph{weak commutativity condition} or simply as \emph{compatibility condition}. For pure states and unitary encodings, the Uhlmann curvature takes the simple form $\mathcal U = \text{Im}\mel{\psi_0}{\mathcal H \mathcal H^T}{\psi_0}$, where $\mathcal H = i\hat U_{\boldsymbol x}^\dagger\grad U_{\boldsymbol x}$ is a vector arranging the hermitian generators associated to each variable. Explicitly, here we have (up to unimportant scalar additions) $\mathcal H = \mqty( \hat p & -\hat q)^T$. From where it follows, 
    \begin{equation}
        \mathcal U = \mqty(0 & 1/2\\ 1/2 & 0),
    \end{equation}
    where we used the fact that $\Im\ev{\hat q \hat p} =1/2$ for any state, something that follows from $[\hat q, \hat p] = i$. Hence, for any finite energy sensor state, the Holevo bound does not saturate the two-quadrature displacement estimation QCRB. However, we know that in the limit of infinite energy it does. The transition is explained by the ratio between the Uhlmann curvature and the quantum Fisher information matrix, which bounds the ratio between the Holevo and quantum Cramer-Rao bounds \cite{carollo2019quantumness}, 
    \begin{equation}
    \label{appendix:compatibility}
        1\leq \frac{\mathcal B_H(\rho_{\boldsymbol x}, \mathsf W)}{\tr(\mathsf W \mathsf F_Q^{-1})}\leq 1+\mathcal R \leq 2,
    \end{equation}
    with $\mathcal R = ||i\mathsf F_Q^{-1}\mathcal U||_{\infty}$, where $||\bullet||_\infty$ indicates the largest eigenvalue norm. Explicitly,
    \begin{equation}
    \label{Appendix:holevo-bound}
        \frac{1}{2\bar n+1} + \frac{1}{4(2\bar n+1)^2} \geq \mathcal B_H(\rho_{\boldsymbol x}, I).
    \end{equation}
    Hence, the Holevo bound approaches the QCRB as $1/16n^2$, and as such, for even moderate number of photons the QCRB can in principle be nearly saturated. In the asymptotic limit, where the Holevo bound is guaranteed to be saturated with an appropriate choice of measurement basis and estimator, the LHS of \cref{Appendix:holevo-bound} upper bounds the achievable precision.

    Going beyond unbiased estimators, if these are biased, the Jacobian matrix of the estimators $\mathsf{B}(\rho_{\boldsymbol x}, \mathcal E, \tilde{\boldsymbol x}) = \text E[\partial \tilde{\boldsymbol x}/\partial \boldsymbol x]$ weights the Fisher information matrix, and instead of \cref{appendix:mQCRB} we have the inequality chain, 
    \begin{equation}
    \label{Appendix:bias}
        \mathsf C \geq \mathsf B \mathsf{F}^{-1} \mathsf B^T\geq \mathsf B \mathsf F_Q^{-1} \mathsf B^T.
    \end{equation}
    Hence, taking the trace, and assuming both quadratures are symmetric, we obtain the $1/\sqrt{4\bar n+2}$ bound on the achievable sensitivity.  In the asymptotic limit, and in virtue of  \cref{Appendix:holevo-inequality,appendix:compatibility,Appendix:holevo-bound,Appendix:bias}, the achievable sensitivity is upper bounded by $\sqrt{1+1/(8\bar n+4)}/\sqrt{4\bar n+2}$.

    
    \section{small-Big-small stabilization protocol}
    \label{Appendix:sBs}

    Here, we provide details on the small-Big-small (sBs) stabilization protocol introduced in Refs. \cite{Royer:2020,DeNeeve:2022}, and used in the main text. First, we establish notation. We denote the bosonic annihilation operator as $\hat a$, quadratures $\hat q = (\hat a^\dagger+\hat a)/\sqrt{2})$, $\hat p = i(\hat a^\dagger-\hat a)/\sqrt{2}$, and modular quadratures $\hat x_{[m]}\equiv \hat x\mod m$, with $x\in \{q,p\}$. Rotated quadratures are denoted as $\hat x_\theta = \hat R(\theta)\hat x \hat R(\theta)^\dagger$, with $\hat R(\theta)=e^{-i\theta\hat n}$, and the displacement operator $\hat D(\beta)= \exp(\beta \hat a^\dagger - \beta^*\hat a)$. We abbreviate hyperbolic functions as $c_\Delta = \cosh(\Delta^2),s_\Delta =\sinh(\Delta^2), t_\Delta=\tanh(\Delta^2)$. We write Pauli matrices as $\sigma_k$ with $k\in\{x,y,z\}$, and adorn all other quantum operators with a hat.
    
    The sBs protocol aims to stabilize the finite-energy qunaught state via trotterization of the Lindblad dynamics,
    \begin{equation}
    \label{eq:Lindblad-ideal}
        \frac{\dot{\hat\rho}}{\Gamma} = D[\hat d_{q,\Delta}]\rho + D[\hat d_{-,\Delta}]\rho \equiv \mathcal L \hat\rho,
    \end{equation}
    where $\hat d_{x,\Delta} \propto \ln \hat T_{x,\Delta} =  (\hat x_{[l/c_\Delta]}/\sqrt{t_\Delta}+i \hat x_{\pi/2}\sqrt{t_\Delta} )$, and $\Gamma$ is the decay rate, to be set below. Hence,~$\mathcal L \ketbra{\#_\Delta}=0$, the finite-energy qunaught state is the steady state. The trotterization is done on the unitaries obtained after modelling the oscillator-bath interaction with an oscillator-qubit one, 
    \begin{equation}
        \hat U^{\text{target}}_{x,\Delta} = \exp(-i\sqrt{\frac{\Gamma\delta t}{2}}[\hat d_{x,\Delta}^\dagger (\sigma_x+i\sigma_y) +\text{h.c.}]),
    \end{equation}
    followed by reset of the qubit after each interaction. Approximating the bath as a qubit is accurate whenever $\sqrt{\Gamma \delta t}$ is small. The next step is to perform the trotterization $e^{\hat A_{x} + \hat B_{x}}\simeq e^{\hat A_{x}/2}e^{\hat B_{x}}e^{\hat A_{x}/2}$. The choice, $\hat A_{x} \propto \hat x_{\pi/2}$ and $\sqrt{\Gamma\delta t} = l\sqrt{c_\Delta s_\Delta}/2$, leads upon rearrangement to the sBs circuit shown in \cref{fig:setup}. The alternative choice, $\hat A_x\propto \hat x_{[l/c_\Delta]}$ leads to the Big-small-Big protocol, analyzed for metrology in Ref.~\cite{valahu2024quantum}. The choice of  $\sqrt{\Gamma\delta t}$ ensures the modularity of the unitary, that is, invariance under modular displacements $e^{i(l/c_\Delta)\hat x_{\pi/2}}\hat U_{x,\Delta}e^{-i(l/c_\Delta)\hat x_{\pi/2}}=-\hat U_{x,\Delta}$. Then, the effect on the stabilizers is approximately given by $\hat U_{x,\Delta} \hat T_{x_{\pi/2},\Delta}\hat U_{x,\Delta}^{\dagger}\simeq -\hat T_{x_{\pi/2},\Delta}$ and $\hat U_{x,\Delta} \hat T_{x,\Delta}\hat U_{x,\Delta}^{\dagger}\simeq \hat T_{x,\Delta}$. In other words, the unitary take the state of the oscillator from the $+1$ to the $-1$ eigenstate of $\hat T_{x_{\pi/2},\Delta}$.  To ensure stability, in the next application we update the definition of $\hat d_{x_{\pi/2},\Delta}\propto \ln \hat T_{x_{\pi/2},\Delta} \to \ln (-\hat T_{x_{\pi/2},\Delta})$. 
    
    To deal with these updates, results convenient to define a gauge vector $\vb j = (j_q,j_p)$ with $j_x \in \{0,1\}$. Taking into account this gauge change, and upon simplifications using the identities $\sigma_x = \hat H \sigma_z \hat H$ and $\sigma_y = \hat R_z(\pi/2)\hat H \sigma_z \hat H\hat R_z^\dagger(\pi/2)$, we obtain the sBs unitaries, 
    \begin{subequations}
    \label{eq:sBs-unitaries}
        \begin{align}
        \begin{split}
            \hat U^{j_q}_{q,\Delta} &= C\hat D\qty(\frac{ls_\Delta}{2})\hat{R}_x^\dagger\qty(\nu_q\frac{\pi}{2})C\hat D\qty(-ilc_\Delta)\\
            &\qq{}\qq{}\qq{}\times \hat{R}_x\qty(\frac{\pi}{2})C\hat D\qty(\frac{ls_\Delta}{2}),
        \end{split}
        \\
        \begin{split}
            \hat U^{j_p}_{p,\Delta} &= C\hat D\qty(i\frac{ls_\Delta}{2})\hat{R}_x^{\dagger}\qty(\nu_p\frac{\pi}{2})C\hat D\qty(lc_\Delta)\\
            &\qq{}\qq{}\qq{}\times \hat{R}_x\qty(\frac{\pi}{2}) C\hat{D}\qty(i\frac{ls_\Delta}{2}),
        \end{split}
        \end{align}
    \end{subequations}
    with $\nu_x = (-1)^{j_x}$, and $C\hat D(\alpha) = \exp[(\alpha \hat{a}^\dagger - \alpha^*\hat{a} )\sigma_z/2\sqrt{2}]$. These unitaries, displayed in circuit form in \cref{fig:setup}, stabilize by turns the four joint $+1$ eigenstates of $\{(-1)^{j_q}\hat T_{q,\Delta},(-1)^{j_p}\hat T_{p,\Delta}\}$. These four eigenstates are nearly translational invariant with respect to one another, hence their metrological power for displacement sensing is equivalent. The finite-energy qunaught state we use is the joint eigenstate for gauge $\vb j = (0,0)$. For more details on the unitaries derivation, see Ref.~\cite{Royer:2020}.
    
    For metrology, the last control displacement is substituted by measurement and feedback. The Kraus operators of each subround $q,p$, defined as $\hat K_{x,g/e} = \bra{g/e}\hat U_{x,\Delta}'\ket{+}$, are
    \begin{equation}
    \begin{split}
    \label{appendix:eq-kraus}
    \hat K_{e/g,x}^{j_x} &= \frac{1}{\sqrt{2}}[e^{-i\varphi}\cos(\frac{lv}{2}\hat x_{\theta_+}\pm\nu_x\pi/4)\\
    &-ie^{i\varphi}\cos(\frac{lv}{2}\hat x_{\theta_-}\pm \nu_x\pi/4)],
    \end{split}
    \end{equation}
    with the $+$ sign of $\pm$ associated to $e$, $\varphi = \pi s_\Delta c_\Delta/8$, $v=\sqrt{c_\Delta^2+s_\Delta^2/4}$, and $\theta_{\pm}=\arctan(\pm t_\Delta/2)$. Hence, the measurement of the qubit is an indirect measurement of the modular quadratures of the oscillator. To second order in $\Delta$, the Kraus operators take the simple form (up to normalization), 
    \begin{equation}
    \label{eq:Kraus-delta2}
    \begin{split}
    \hat K_{e/g,x}^{j_x} &\propto \cos(\frac{lc_\Delta}{2}\hat x\pm \nu\pi/4)\\
    &-i\sin(\frac{lc_\Delta}{2}\hat x\pm \nu\pi/4)\sin(\frac{ls_\Delta}{4}\hat x_{\pi/2}).
    \end{split}
    \end{equation}
    The first term, the ideal one, is a modular measurement of the $x$ quadrature. The second term is a weak modular measurement of both quadratures simultaneously, reduces the contrast obtained by the first term, and is responsible for the stabilization. Additionally, using \cref{eq:Kraus-delta2}, a simple calculation shows that the measurement probabilities do not depend on $\hat x_{\pi/2}$ up to fourth order in $\Delta$. Then, the main backaction in the opposite quadrature is via the feedback. When taking the average, the action of the feedback cancel each other, which explains the observed independency of the quadratures on average. 
    
\begin{figure}[t]
        \centering
        \includegraphics[width=\linewidth]{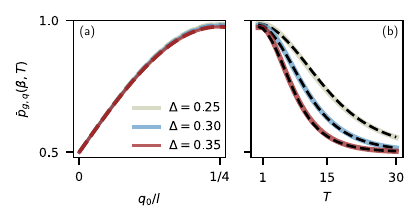}
        \caption{Probabilities of measuring the qubit in the ground state in the first subround a), and after several rounds using the averaged recovery map b). a) As a function of $q_0$ with $p_0=l/4$ fix, after a displacement $\hat D(\beta)$ with $\beta=(q_0+ip_0)/\sqrt{2}$ on the ideal sensor state $\ket{\#_\Delta}$. Solid lines are numerics, and dashed lines correspond to \cref{eq:p_eg_x_single} with $a_1=0.4$. b) Evolution of the probability as a function of the number of rounds $T$, following an initial displacement with $ q_0= p_0=l/4$. Solid lines are numerics. Dark dashed lines are \cref{eq:pex-improved}, with $a_2=1.44$, $a_3=0.44$, fitted for $\Delta=0.3$. The same $a_2$ and $a_3$ are used for all three values of $\Delta$.}
        \label{fig:pg-singleround-severalrounds}
    \end{figure}

    Concretely, for a single subround, i.e. $\rho \to \hat K_{e/g}^{j_x}\rho (K_{e/g}^{j_x})^\dagger$, the probabilities of measuring the qubit in $e/g$ are, 
    \begin{equation}
        p^{j_x}_{e/g,x} \simeq \frac{1}{2}[1\mp \nu_x C_K(\Delta)\langle\sin(lc_\Delta \hat x)\rangle],
    \end{equation}
    with $C_K(\Delta)$ a contrast function. The expectation value $\langle\sin(lc_\Delta \hat x)\rangle$ can be approximated analytically. However, given that we also need to account for the contrast $C_K(\Delta)$, we find it convenient to express the probabilities as, 
    \begin{equation}
    \label{eq:p_eg_x_single}
        p^{j_x}_{e/g,x} = \frac{1}{2}[1\mp \nu_x e^{-a_1\Delta^2}\sin(lc_\Delta \langle\hat x\rangle)].
    \end{equation}
    In \cref{fig:pg-singleround-severalrounds}~a) we plot these probabilities for $\Delta=0.3$, and displacements $\hat D(q_0\sqrt{2})$ with $q_0$ $\in [0,l/4]$. We find $a_1=0.4$ to be an excellent fit. For numerical calculations, we use the fock basis with dimension $n=150$. All codes are available at \cite{Repository_for_Quantum_2025}.

    Going beyond the single subround case, to gain intuition we focus on the probabilities after the application of the average recovery map, that is $\rho \to \mathcal R_T \rho $. Heuristically, to stabilize a grid with finite energy, there must be a drift to the center of phase space, along a drift towards each point in the lattice. Hence, when the sensor state is displaced within the first lattice cell, its center, $(\langle \hat q\rangle , \langle\hat p\rangle)$, feels an attraction both towards the center of phase space, and to the nearest grid point. The simple expression for the probability of measuring the qubit \cref{eq:p_xg}, captures the attraction felt towards the center of phase space. The following, 
    \begin{equation}
    \label{eq:pex-improved}
        \bar p_{e/g,x}(\beta,T) = \frac{1}{2}\mp\frac{\nu_x}{2}e^{-a_1\Delta^2}\sin\bigl(lc_\Delta x_0e^{-f(x_0,T)\Delta^2(T-1)}\bigr),
    \end{equation}
    with $f(x_0,T) = a_2-a_3|\sin(lx[T])|$, and $x[T]=q_0e^{-f(x_0,T-1)\Delta^2(T-1)}$,    
    captures both. In \cref{fig:pg-singleround-severalrounds} b), we plot $p_{e,q}$ up to 30 rounds for $\Delta\in \{0.25,0.3,0.35\}$. We find excellent agreement between \cref{eq:pex-improved} and the numerically obtained probabilities.  

    \section{Noiseless metrological performance of the sBs protocol}
    \label{appendix:noiseless}
    
    In this appendix, we provide further details on the metrological performance of the protocol presented in the main. We discuss our simulation methods, and provide a detailed discussion on how we obtained the estimators. For flat priors $q_0,p_0\in \{-l/4,l/4\} $, we provide results on the sensitivity achieved as a function of $\Delta$, and compare it with the bound obtained in \cref{appendix:quantum-bound}. When Gaussian priors are used, we provide the mean-square error, and its variance, as a function of $\Delta$ and number of rounds of estimation. 
    Note that it is necessary to discretize the space of possible displacements, and we numerically confirm the convergence of the chosen discretization. Throughout this appendix, for ease of notation we omit the gauge choices and the small feedback displacements, both included in the simulations. All codes used to obtain the results shown are available at~\cite{Repository_for_Quantum_2025}. 
    
    \subsection{Numerical construction of the estimators}
    
    We do our simulations in the fock basis, with a Hilbert space dimension of 140 for  $\Delta \in \{0.25,0.4\}$. The initial state we use, $\rho_0$, is the steady state of the sBs protocol, that we numerically obtained after 120 rounds of autonomous sBs, with the vacuum as initial state. To be concrete, one round of autonomous sBs consists of $\rho_{\text{in}} \to \rho_{q} = \hat K_{g,q}\rho_{\text{in}} \hat K_{g,q}^\dagger + \hat K_{e,q}\rho_{\text{in}} \hat K_{e,q}^\dagger \to \rho_{\text{out}} = \hat K_{g,p}\rho_{q} \hat K_{g,p}^\dagger + \hat K_{e,p}\rho_{q} \hat K_{e,p}^\dagger$. We numerically obtain the Kraus operators from their definition $\hat K_{x,g/e} = \bra{g/e}\hat U_{x,\Delta}'\ket{+}$. The resulting state is nearly exactly the qunaught GKP state with associated gauge vector $\boldsymbol j = (0,0)$. We check this is the case, by computing the expectation value of the stabilizers $\hat T_{x,\Delta}$ which we found to be close to 1 up to 3 decimal points.

    The estimators are obtained from the probabilities of measuring the qubit bitstring $\boldsymbol b$, that consists of $2T$ bits after $T$ rounds of sBs. Due to the near independency of the quadratures, we split these bitstrings in two bitstrings of length $T$, one for each quadrature, and write $\boldsymbol b = \mqty( \boldsymbol b_q & \boldsymbol b_p)$. We obtain the estimator of the initial displacement $q_0$ $(p_0)$, from the resulting bitstring $\boldsymbol b_q$, and we write it as $\tilde q_0(\boldsymbol b_q)$ ($\tilde p_0(\boldsymbol b_p)$). The maximum likelihood estimator is defined as the value of $q_0$ that maximizes the distribution $p(q_0|\boldsymbol b_q) \propto p(\boldsymbol b_q|q_0)p(q_0)$, where $p(q_0)$ is the prior distribution. The Bayesian estimator we use, is the mean value of the posterior distribution $p(q_0|\boldsymbol b_q)$, that is
    \begin{equation}
        \tilde q_0(\boldsymbol b_q)_B = \int \dd q_0 p(q_0) p(q_0|\boldsymbol b_q),
    \end{equation}
    where $p(q_0) = e^{-q_0^2/2\sigma^2}/\sqrt{2\pi\sigma^2}$ is the prior distribution. As already mentioned in the main, this choice is so as it minimizes the averaged over the prior mean-square error $\int \dd q_0 p(q_0)\Ev[(\tilde q_0(\boldsymbol b_q)- q_0)^2]_{q_0}$, where $\Ev[(\tilde q_0(\boldsymbol b_q)- q_0)^2]_{q_0} = \sum_{\boldsymbol b_q}p(\boldsymbol b_q|q_0)[\tilde q_0(\boldsymbol b_q)-q_0]^2$ is the average taken over the possible bitstrings for a given displacement $q_0$. The same is done for the $p$ quadrature, now with $\boldsymbol b_p$. Hence, to build the estimators we need the probabilities of measuring the bitstrings $\boldsymbol b_q$, $\boldsymbol b_p$, for a grid of values $q_0, p_0$, with large enough range and fine graining to assure convergence in the computation of the maximum and mean of the posterior distributions. 

    By definition, $p(\boldsymbol b_q|q_0) = \int \dd p_0 \sum_{\boldsymbol b} p(\boldsymbol b|q_0,p_0)$, with $\boldsymbol b_q$ in $\boldsymbol b$. Each one of the probabilities $p(\boldsymbol b|q_0,p_0)$ can be computed exactly as $\tr\{\hat K_{\boldsymbol b}\rho_0(q_0,p_0)\hat K_{\boldsymbol b}^\dagger\}$, where $\hat K_{\vb{b}} = \prod_{t=T}^1 \hat K_{b_t^p,p}\hat K_{b_t^q, q}$, and $\rho_0(q_0,p_0)= \hat D(\beta)\rho_0\hat D(\beta)^\dagger$ with $\beta = (q_0+ip_0)/\sqrt{2}$.
    Clearly, carrying out this computation is prohibitive for a large $T$, e.g. for $T=12$ a total of $2^{24}$ probabilities would need to be computed. Conveniently, the near independence between the quadratures greatly simplifies the problem. Concretely, this independence translates in the fact that the probability of measuring the bitstring $\boldsymbol b_q$, is nearly independent of both the initial displacement in the $p$ quadrature $p_0$, and the probability of measuring $\boldsymbol b_p$. In mathematical terms, $p(\boldsymbol b_q,\boldsymbol b_p|q_0,p_0) \simeq p(\boldsymbol b_q,\boldsymbol b_p'|q_0,p_0')$ for any pair of $p$ quadrature bitstrings  $(\boldsymbol b_p,\boldsymbol b_p')$, displacements in the $p$ quadrature ($p_0$, $p_0'$), $q$ quadrature bitstring $\boldsymbol b_q$, and initial displacement $\hat D(\beta)$. Hence, we can use the probabilities $p(\boldsymbol b_q|q_0,0)$ ($p(\boldsymbol b_p|0,p_0)$) to estimate the displacement in $q$ ($p$) quadrature $q_0$ ($p_0$), with minimal losses in performance. We compute $p(\boldsymbol b_q|q_0,0)$ using the update rule $\rho_{t}(\beta) \to \rho_{t,q}(\beta) = \hat K_{b_q^t,q}\rho_t(\beta) \hat K_{b_q^t,q}^\dagger\to \rho_{t+1}(\beta) = \hat K_{g,p}\rho_{t,q}(\beta) \hat K_{g,p}^\dagger + \hat K_{e,p}\rho_{t,q}(\beta) \hat K_{e,p}^\dagger$, with $\beta=q_0/\sqrt{2}$, and  $p(\boldsymbol b_q|q_0,0)= \tr[\rho(\boldsymbol b_q,q_0)]$, where $\rho(\boldsymbol b_q,q_0)$ is the unnormalized density matrix obtained after the application of the update rule for all bits in $\boldsymbol b_q$. This allows us to efficiently and exactly compute these probabilities.  To compute the probabilities of measuring $\boldsymbol b_p$ given $p_0$, we use the same approach, now with update rule $\rho_{t}(\beta) \to \rho_{t,q}(\beta) = \hat K_{g,q}\rho_t(\beta) \hat K_{g,q}^\dagger+\hat K_{e,q}\rho_t(\beta) \hat K_{e,q}^\dagger\to \rho_{t+1}(\beta) = \hat K_{b_p^t,p}\rho_{t,q}(\beta) \hat K_{b_p^t,p}^\dagger$, and $\beta=ip_0/\sqrt{2}$.


    Once the probabilities of measuring each bitstring $\boldsymbol b_q$ after $T$ rounds of estimation are obtained, building the estimators of the initial displacement $q_0$ ($p_0$) for each $q$ ($p$) bitstring $\tilde q_0(\boldsymbol b_q)$ ($\tilde p_0(\boldsymbol b_p)$), is straightforward. For clarity, we summarize the steps we do to obtain the estimators in list form, 
    \begin{itemize}
        \item Step 1. The initial sensor state $\rho_0$, is displaced to $\rho_0(\beta)= \hat D(\beta)\rho_0 \hat D(\beta)^\dagger$. Repeat this step and all that follow for different values of $q_0$, $p_0$. To check the discretization chosen, ensure convergence of the resulting estimators.
        \item Step 2. To compute $p(\boldsymbol b_q|q_0)$ apply the unravelled in $q$ sBs update $\rho_{t}(\beta) \to \rho_{t,q}(\beta) = \hat K_{b_q^t,q}\rho_t(\beta) \hat K_{b_q^t,q}^\dagger\to \rho_{t+1}(\beta) = \hat K_{g,p}\rho_{t,q} \hat K_{g,p}^\dagger + \hat K_{e,p}\rho_{t,q} \hat K_{e,p}^\dagger$, with $\beta=q_0/\sqrt{2}$. Then, the probability is given by $ p(\boldsymbol b_q|q_0) = \tr[\rho(\boldsymbol b_q, q_0)]$, and repeat for all $\boldsymbol b_q \in \mathbf Z_2^T$. Do a similar computation for the $p$ quadrature, now with update rule $\rho_{t}(\beta) \to \rho_{t,q}(\beta) = \hat K_{g,q}\rho_t(\beta) \hat K_{g,q}^\dagger+\hat K_{e,q}\rho_t(\beta) \hat K_{e,q}^\dagger\to \rho_{t+1}(\beta) = \hat K_{b_p^t,p}\rho_{t,q}(\beta) \hat K_{b_p^t,p}^\dagger$, and $\beta=ip_0/\sqrt{2}$. 
        \item Step 3. Construct the estimators of the initial displacements associated to each bitstring, $\tilde q_0(\boldsymbol b_q)$, $\tilde p_0(\boldsymbol b_p)$. The maximum likelihood estimator is the value of $\tilde q_0\in [-l/4,l/4]$ where the maximum value of $\tilde p(\boldsymbol b_q|q_0)$ is reached. The Bayesian estimator is the mean of the posterior $\tilde p(q_0|\boldsymbol b_q) \propto \tilde p (\boldsymbol b_q|q_0)p(q_0)$, with Gaussian prior $p(q_0) = e^{-q_0^2/2\sigma}/\sqrt{2\pi \sigma^2}$. Numerically, we set $q_0,p_0\in [-l,l]$, a range that allows us to accurately compute the aforementioned mean.
    \end{itemize}

    \begin{figure}
        \centering
        \includegraphics[width=\linewidth]{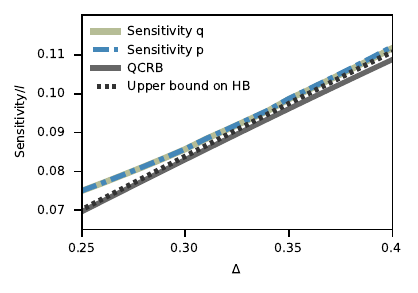}
        \caption{Noiseless sensitivity at $q_0=p_0=0$ as a function of $\Delta$. Solid red (dashed blue) line is the sensitivity achieved in the estimation of the $q$ ($p$) quadrature displacement $q_0$ ($p_0$). Solid black line is the bound on the sensitivity given by multivariate quantum Cramer-Rao bound (QCRB). Dotted black line is the upper bound on the Holevo bound derived in \cref{appendix:quantum-bound}.}
        \label{fig:sensitivity-bounds-delta}
    \end{figure}

    \subsection{Metrological performance}
    \label{appendix:metrological-performance}
    
\begin{figure}[t]
        \centering
        \includegraphics[width=\linewidth]{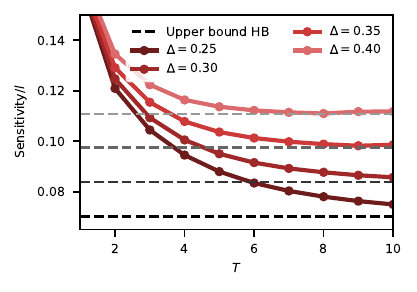}
        \caption{Sensitivity in the estimation of the $q$ quadrature displacement at $q_0=p_0=0$, as a function of the number of bits acquired, for different envelopes. Dashed lines correspond to the upper bound on the Holevo bound derived in \cref{appendix:quantum-bound}.}
        \label{fig:sensitivity-convergence}
    \end{figure}

     First, we focus on the case where we expect the displacement to be small, and we use a flat prior $q_0,p_0 \in [-l/4,l/4]$. In \cref{fig:sensitivity-bounds-delta}, we plot the sensitivity achieved in the estimation of each of the quadratures at $q_0=p_0=0$, after $T=10$ rounds of estimation, as a function of the envelope width $\Delta$. The sensitivity achieved in the estimation of $q_0$ (solid red line), is nearly equal to the one achieved in the estimation of $p_0$ (dashed blue line). Both of them, approach the upper bound on the Holevo bound (dotted black line), given by $\sqrt{1+1/(8\bar n+4)}/\sqrt{4\bar n+2}$. In the figure, as $\Delta$ decreases the distance to the bound increases. This is explained by the fact that for $T=10$, not all the available information of the displacement has been extracted for these lower values of $\Delta$. This gap is better understood looking at \cref{fig:sensitivity-convergence}. There, we plot the sensitivity achieved on the estimation of $q$ quadrature, at $q_0=p_0=0$, as a function of the number of bits acquired for different envelopes (solid lines). For the smaller envelope (larger $\Delta$), the upper bound on the Holevo bound (dashed lines) is nearly saturated, and as more bits are acquired the performance does not improve. For a larger envelope with $\Delta=0.25$, up to $T=10$ bits acquired the decaying slope suggests that as with the other envelopes, the bound will be approached if more bits are acquired.
    
    In our simulations, for each value of $\Delta$ in \cref{fig:sensitivity-bounds-delta}, we used a range of $q_0\in [-l/4,l/4]$ ($p_0\in [-l/4,l/4]$) with $51$ values on it, which translates on 52224 probabilities computed for each point in the plot. Thus, we note that a path to improve the efficiency of the simulations, reach higher values of $T$ and envelopes with smaller $\Delta$, will be using the sBs basis developed in Ref.~\cite{Hopfmueller2024bosonicpauli}, instead of the fock basis as we do here.

\begin{figure}
        \centering
        \includegraphics[width=1\linewidth]{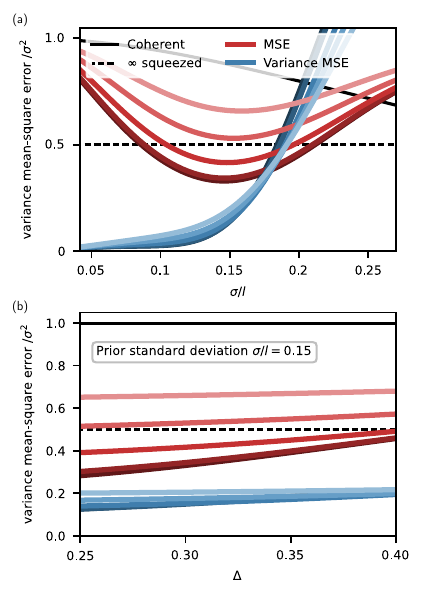}
        \caption{Mean-square error and variance of the mean-square error when estimating the displacement in the $q$ quadrature $q_0$. a) As a function of the prior standard deviation, with fix envelope width $\Delta=0.3$. b) As a function of the envelope width $\Delta$, with a fix prior standard deviation of $\sigma/l = 0.15$. Lighter to darker lines correspond to the number of acquired bits per quadrature $T\in [1,2,4,8,10]$. Solid (dashed) black line is the coherent (Gaussian) limit of displacement estimation.}
        \label{fig:variance-gaussian}
    \end{figure}

    When doing estimation with Gaussian priors $p(q_0,p_0)=\mathcal G_\sigma(q_0)\mathcal G_\sigma(p_0)$, the main figure of merit we use is the averaged over the prior mean-square error,
    \begin{equation}
        \text{E}[(\tilde q_0(\boldsymbol b_q)-q_0)^2] = \int \dd q_0 \mathcal G_\sigma(q_0)\text{E}[\boldsymbol(\tilde q_0(\boldsymbol b_q)-q_0\boldsymbol)^2]_{q_0}, 
    \end{equation}
    where $\text{E}[\boldsymbol(\tilde q_0(\boldsymbol b_q)-q_0\boldsymbol)^2]_{q_0} = \sum_{\boldsymbol b_q}p(\boldsymbol b_q|q_0)\text{E}\boldsymbol(\tilde q_0(\boldsymbol b_q)-q_0\boldsymbol)^2$.
    A second relevant quantity is the variance of this mean-square error defined as $\Var[(\tilde q_0(\boldsymbol b_q)-q_0)^2]_{q_0} = \text{E}\{\boldsymbol(\tilde q_0(\boldsymbol b_q)-q_0\boldsymbol)^2-\text{E}[\boldsymbol(\tilde q_0(\boldsymbol b_q)-q_0\boldsymbol)^2]\}_{q_0}$.
    This is so because, in the finite sample regime, the averaged error will differ from the theoretical one, which assumes an infinite number of samples. Hence, the averaged over the prior variance of the mean-square error $\int \dd q_0 p(q_0)\Var[(\tilde q_0(\boldsymbol b_q)-q_0)^2]_{q_0}$, is an indicator of how much deviation from the asymptotic mean to expect. In \cref{fig:variance-gaussian} a), we plot the mean-square error, and its variance, as a function of the prior standard deviation, for $\Delta=0.3$. Around $\sigma/l=0.15$, where the minimal ratio between the mean-square error and the prior is achieved, the variance is around a third of the mean-square error. This relatively large variance explains the fluctuations in performance shown in \cref{fig:backaction}. In \cref{fig:variance-gaussian} b), we plot the value of the mean-square error and its variance at $\sigma=0.15l$ for different $\Delta$. As  $\Delta$ decreases, so do the mean-square error and its variance. The improvement in performance is approximately linear in $\Delta$ independent of the number of bits acquired. Due to the slow down of the stabilization as $\Delta$ increases, the slope is steeper when more bits are acquired. Additionally, the ratio between the variance and the mean-square error, decreases slowly as $\Delta$ decreases, going from $0.32$ at $\Delta=0.4$ to $0.31$ at $\Delta=0.25$. In the estimation of the $p$ quadrature displacement $p_0$, similar performances are achieved.

    \subsection{Backaction evading protocol.}
    \label{appendix:backaction}
    
\begin{figure}[t]
        \centering
        \includegraphics[width=\linewidth]{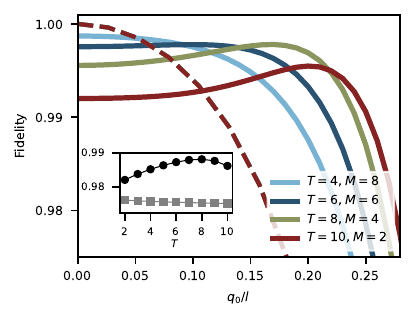}
        \caption{Averaged recovery fidelity as a function of the initial displacement $q_0$, with Bayesian estimation and Gaussian prior with standard deviation $\sigma = 0.15l$, for different $T$, $M$ values of the backaction evading protocol. Solid (dashed) lines are the achieved fidelity with (without) the recovery displacement.
        We use an envelope with $\Delta=0.3$. The dark (light) markers in the inset show the weighted by the prior $p(q_0)=\mathcal G_\sigma(q_0)$ averaged recovery fidelity with (without) the recovery displacement. See text for a detailed explanation. }
        \label{fig:avg-recovery-fidelity}
    \end{figure}

    We characterize the backaction evading performance as follows. First, we determine the averaged recovery fidelity as a function of the initial displacement in the $q$ quadrature, see \cref{fig:avg-recovery-fidelity}. The recovery fidelity is $\mathcal F[\rho_R(q_0,\boldsymbol b,M), \rho_{\#_\Delta}] = \tr[(\sqrt{\rho_R(q_0,\boldsymbol b,M)}\rho_{\#_\Delta}\sqrt{\rho_R(q_0,\boldsymbol b, M})^{1/2}]$, where  $\rho_R(q_0,\boldsymbol b, M)$ is the cavity state after the backaction evading protocol, as described in the main text. Then, the average is $\bar{\mathcal F}[\rho_R(q_0,\boldsymbol b,M), \rho_{\#_\Delta}] =\sum_{\boldsymbol b_q} p(\boldsymbol b_q|q_0)\mathcal F[\rho_R(q_0,\boldsymbol b,M), \rho_{\#_\Delta}]$.
    
    In \cref{fig:avg-recovery-fidelity}, we set the total bit budget to 12, the standard deviation of the Gaussian prior to be $\sigma=0.15l$, and change the number of bits used in the sBs metrology protocol $T$ (solid lines). We also plot the achieved fidelity without the recovery displacement (dashed lines). Only one dashed line is perceivable as for all other values of $T$, $M$ nearly the same fidelity is achieved. For larger initial displacements $q_0$, the recovery displacement improves the fidelity. This is due to the improvement in the accuracy estimation. However, for small values of $q_0$ we see that as $T$ increases the recovery fidelity near $q_0=0$ decreases. This is due to the unavoidable estimation error, and the smaller values of $M$. To choose $T$ and $M$, we compute the weighted averaged recovery fidelity $\int \dd q_0 p(q_0)\bar{\mathcal F}[\rho_R(q_0,\boldsymbol b,M), \rho_{\#_\Delta}]$ where $p(q_0) = e^{-q_0^2/2\sigma}/\sqrt{2\pi\sigma^2}$ is the prior distribution. We find that for $T \in [2,10]$, this weighted fidelity is maximized at $T=8$, see inset in \cref{fig:avg-recovery-fidelity}. Additionally, in the inset we show the weighted averaged recovery fidelity without the recovery displacement, nearly constant for all values of $T$ and $M$, and substantially smaller than the one achieved with the recovery displacement. 

    \section{Noise analysis details}
    \label{appendix-noise}

    In this appendix, we provide further results in the performance of the sBs metrology protocol in presence of noise. We provide the fidelity of the sBs steady state for different noise strengths, as a function of the envelope width $\sim 1/\Delta$. We compare the performance in the estimation of both $q$ and $p$ quadrature displacements. We also provide the scaling of the performance as a function of each type of noise, i.e. relaxation and dephasing of the qubit and cavity, and identify which one is most detrimental. Additionally, we show the backaction evading performance in the presence of noise. We stress that throughout this appendix we focus on qualitative features of the performance in the presence of noise. This allows us to identify relevant parameter optimization directions. However, we do not pursue these optimizations here. All our codes are available at~\cite{Repository_for_Quantum_2025}.

    \subsection{Effect of noise in state preparation and choice of envelope width}

\begin{figure}[t]
        \centering
        \includegraphics[width=\linewidth]{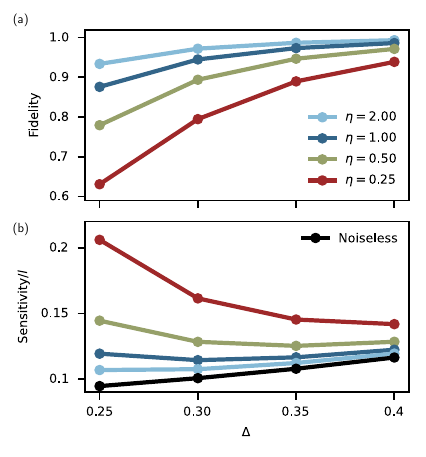}
        \caption{Fidelity of the sBs steady state with the finite-energy qunaught state, and sensitivity achieved in the estimation of the $q$ quadrature at $q_0=p_0=0$ in the presence of noise, after acquiring 4 bits per quadrature, as a function of the envelope width parameter $\Delta$.}
        \label{fig:fidelity-delta-noise}
    \end{figure}

    The initial state we used to obtain the results shown in \cref{fig:noise-performance}, is the steady state of the sBs protocol in presence of noise $\rho_{\#_\Delta,\eta}$, where $\eta$ quantifies the strength of the noise used, as explained in \cref{sec:noise} of the main. To numerically obtain this state, we start the cavity in the vacuum state and run 100 rounds of autonomous sBs stabilization in the presence of noise. In \cref{fig:fidelity-delta-noise} (top), we plot the fidelity of the resulting state with the finite-energy qunaught state for different envelopes and noise levels. For an envelope with $\Delta=0.3$, and $\eta\in[0.25,0.5,1.0,2.0]$, the resulting fidelities are $[0.79, 0.89, 0.94, 0.97]$. The corresponding achieved sensitivities at $q_0=p_0$, after acquiring 8 bits per quadrature, for an envelope with $\Delta=0.3$. are the ones achieved with two-mode squeezed states with squeezing degrees in dB of $[8.3, 10.3, 11.6, 12.1]$, as shown for $\eta \in [0.5,1.0,2.0]$ in \cref{fig:noise-performance} of the main text.

    Due to the preparation fidelity varying as a function of the noise and envelope, for each pair $\{T,\eta\}$, there is an optimal envelope width. In \cref{fig:fidelity-delta-noise} (bottom), we plot the achieved sensitivity in the $q$ quadrature estimation at $q_0=p_0=0$, after acquiring 4 bits per quadrature, for $\eta \in \{0.25, 0.5, 1.0, 2.0\}$ and $\Delta \in \{0.25,0.3,0.35,0.4\}$. The strongest the noise, the smaller the envelope width $\sim 1/\Delta$ maximizing the fidelity. Hence, in practice, once the noise is characterized, care should be taken in choosing an optimal pair $\{\Delta, T\}$. 

    \subsection{Effect of imperfect state preparation vs noise during bit acquisition}
    \label{appendix-noise:state-vs-metrology}

    Here, we separate the effects on the performance due to imperfect state preparation, from the noise during the metrology protocol itself. To do so, we compute the performance starting from the sBs steady state in presence of noise $\rho_{\#_\Delta,\eta}$ followed by noiseless sBs bit acquisition, and compare that with the performance obtained from starting the protocol in the finite-energy qunaught state $\rho_{\#_\Delta}$ followed by noisy sBs bit acquisition.

\begin{figure}[t]
        \centering
        \includegraphics[width=\linewidth]{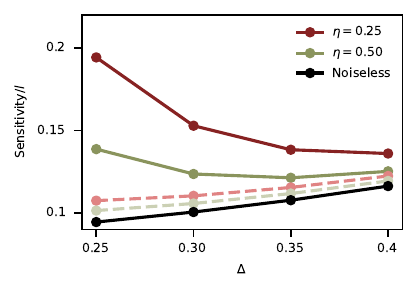}
        \caption{Sensitivity achieved in the estimation of the $q$ quadrature at $q_0=p_0=0$, starting from the noisy sBs steady state followed by noiseless bit acquisition (full colored lines), and starting from the finite-energy qunaught state followed by noisy bit acquisition (dashed lines), after acquiring 4 bits per quadrature, as a function of the envelope width. As comparison, we plot the noiseless sensitivity (dark full line).}. 
        \label{fig:noisystate-vs-idealstate}
    \end{figure}

    Doing so, in \cref{fig:noisystate-vs-idealstate}, we plot the sensitivity in the $q$ quadrature estimation at $q_0=p_0=0$, after acquiring 4 bits per quadrature, as a function of the envelope width, for noise levels set by $\eta \in [0.25,0.5]$. Full lines (dashed lines) are obtained by starting the sBs metrology protocol in the sBs noisy steady state $\rho_{\#_\Delta,\eta}$ (finite-energy qunaught state $\rho_{\#_\Delta}$), followed by noiseless (noisy) acquisition of 4 bits per quadrature. Dark solid line is the noiseless performance. The effect of imperfect state preparation dominates the loss of performance, while the effect of noise during the acquisition of bits induces only a small loss of performance. Hence, using other state preparation methods different from sBs achieving higher fidelity with the finite-energy qunaught state, is the most straightforward pathway to improve the performance shown in \cref{fig:noise-performance}.

    \subsection{$q$ vs $p$ quadrature metrological performance}
    
    The performance achieved in the estimation of both quadratures, as in the noiseless setting, remains practically identical between one another. This is due to the small effect noise has during a single $q$ ($p$) sBs round, as shown by the marginal decrease in performance when noise is only considered in the bit acquisition stage, see \cref{fig:noisystate-vs-idealstate} and discussion in \cref{appendix-noise:state-vs-metrology}. In \cref{fig:noise-q-vs-p}, we plot the sensitivity and mean-square errors achieved in the estimation of each one of the quadratures, for $T\in [1,2,4]$, noise strength set by $\eta = 1.0$, and envelope with $\Delta=0.3$. The performances achieved are practically identical. We find similar results when more bits per quadrature are acquired, different envelope widths, and noise strengths. Naturally, when the strength of the noise increases, the difference slightly increases, but for the largest noise strength we test set by $\eta = 0.25$ we find it remains marginal.
    
    To compute the sensitivities in the estimation of the $q$ ($p$) quadrature, we fix $p_0=0$ ($q_0=0$), and compute the mean-square error for 60 values of $q_0 \in [-l/4,l/4]$ ($p_0 \in [-l/4,l/4]$). This allows us to compute the derivative in the $q_0$ ($p_0$) direction and obtain the sensitivities. To test the accuracy of this approach we repeat the calculation setting $p_0\in [l/4,l/2]$ ($q_0\in[ l/4,l/2]$) and we find nearly identical results.

    \begin{figure}[t]
        \centering
        \includegraphics[width=\linewidth]{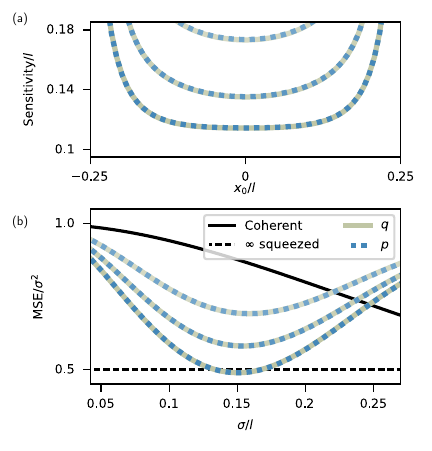}
        \caption{Comparison of metrological performance in the estimation of the $q$ and $p$ quadrature displacements in the presence of noise. a) Sensitivities achieved with maximum-likelihood estimation assuming a flat-prior $x_0\in[-l/4,l/4]$ with $x\in \{q,p\}$. b) Mean-square error achieved with Bayesian estimation assuming a Gaussian prior $p(q_0,p_0)=\mathcal G_{\sigma}(q_0)\mathcal G_\sigma(p_0)$. Lighter to darker lines represent the performance obtained after acquiring 1, 2, and 4 bits per quadrature. The noise strength is set by $\eta=1.0$. The envelope width is set with $\Delta=0.3$.}
        \label{fig:noise-q-vs-p}
    \end{figure}

    \subsection{Effect of each noise}
    \label{appendix:effect-of-each-noise}
    \begin{figure}[t]
        \centering
        \includegraphics[width=\linewidth]{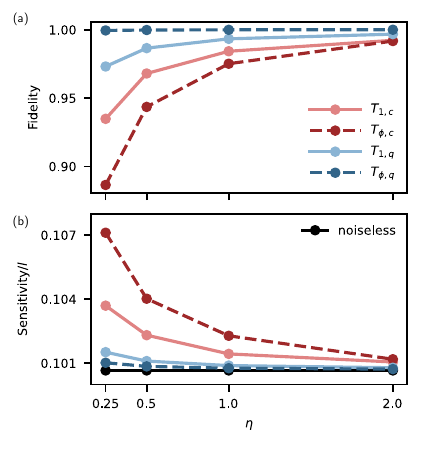}
        \caption{Effect on the performance for each type of noise. Red (blue) lines and dots are cavity (qubit) noise. Solid (dashed) lines are relaxation (dephasing). a) Fidelity with the finite-energy qunaught state of the sBs steady state, as a function of the noise strength set by $\eta$. b) Sensitivity after 4 rounds of the sBs metrology protocol starting from the finite-energy qunaught state. We set an envelope with $\Delta=0.3$.}
        \label{fig:fidelities-sensitivity-eachnoise}
    \end{figure}

    We isolate the effect of each noise, relaxation and dephasing of the cavity (qubit). Our goal is to determine which noise is most detrimental for the performance of the protocol. First, we compute the fidelity of the sBs steady state for each type of noise, setting all other noises to 0, fixing an envelope with $\Delta=0.3$. As done in the main, we set proportionality factors of $\eta \in [0.25,0.5,1.0,2.0]$. In \cref{fig:fidelities-sensitivity-eachnoise}~a), we plot these fidelities, and find that the most detrimental noise is cavity dephasing. A notable feature is that qubit decay is not as detrimental as one may expect. This is so because, in the sBs stabilization of the qunaught state when a bit relaxation event occurs during the large displacement, the resulting state has a large overlap with the one where there was no error. We proceed to explain this last point in details. 
    
\begin{figure}[t]
        \centering
        \includegraphics[width=.9\linewidth]{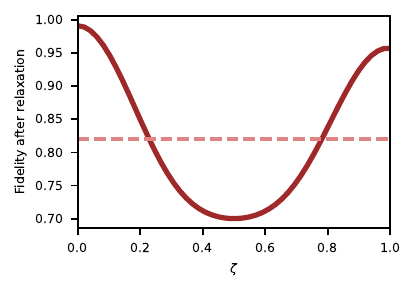}
        \caption{Fidelity between the error-free state after a $q$ quadrature round of sBs, with the resulting state after a qubit relaxation event occurs. The dashed line is the average over $\zeta$ fidelity, approximately equal to 0.82. See text for details.}
        \label{fig:qubit-relaxation}
    \end{figure}

    So far we modelled noise with the Lindblad master equation. Here, to gain intuition, we model the bit relaxation during the large control displacement by the following, 
    \begin{equation}
        C\hat (-ilc_\Delta)\to C\hat (-i\zeta lc_\Delta)\hat P_{g}C\hat (-i(1-\zeta)lc_\Delta),
    \end{equation}
    where $\zeta\in [0,1]$ sets the moment where the qubit relaxes to the ground state. We model this relaxation event by applying the projector onto the ground state $\hat P_g = \ketbra{g}{g}+\ketbra{g}{e}$. We substitute this relaxation included controlled displacement in the definition of the sBs unitaries \cref{eq:sBs-unitaries}, with corresponding Kraus operator $\hat K_{g,\zeta}$. Then, the state after a relaxation event of the qubit in the large displacement of the $q$ quadrature stabilization of $g$ is $\hat \rho_{\zeta} = \hat K_{g,\zeta}\hat \rho_{\#_\Delta}\hat K_{g,\zeta}^\dagger/\text{tr}(\hat K_{g,\zeta}\hat \rho_{\#_\Delta}\hat K_{g,\zeta}^\dagger)$. In \cref{fig:qubit-relaxation} we plot the fidelity of this state with the error-free one $\hat \rho =\hat K_{g/e,x}\hat \rho_{\#_\Delta}\hat K_{g/e,x}^\dagger/\text{tr}(\hat K_{g/e,x}\hat \rho_{\#_\Delta}\hat K_{g/e,x}^\dagger)$, as a function of $\zeta$. We find that for both $g/e$ the fidelity is the same. For $\zeta=0$ the relaxation occurs right at the end of the control displacement, and will affect the measurement statistics but only very mildly the state stabilization. For $\zeta=1$ the relaxation occurs at the start of the controlled displacement, and the resulting effect is that instead of creating a superposition, the cavity is displaced by $\hat D(-ilc_\Delta/\sqrt{2})$. Thus, a qubit relaxation effect has a strong backaction effect on the cavity state. However, due to the high fidelity even if a relaxation event occurs, this error is not as detrimental as in the stabilization of single mode GKP qubits, where relaxation errors can induce logical errors \cite{Royer:2022}.

    Additionally, we explore the effect of each type of noise during the sBs bit acquisition stage. To do so, we compute the sensitivities achieved starting from the finite-energy qunaught state after gathering 4 bits per quadrature, for each type of noise. We plot these sensitivities in \cref{fig:fidelities-sensitivity-eachnoise}~b), setting an envelope with $\Delta=0.3$. The resulting sensitivities are close to the one achieved in the absence of noise. This is due to state preparation being the main source of performance loss, see \cref{appendix-noise:state-vs-metrology}. As with the fidelities, we find cavity decoherence is the most detrimental. However, we note that the gap in the performance between qubit relaxation and dephasing decreases in comparison to the state preparation fidelity. This is due to the fact that both relaxation and dephasing introduced entropy in the qubit measurement statistics, although in different forms. Relaxation introduces a bias toward the ground state, whereas dephasing randomizes the probabilities making them closer to 1/2. Moreover, due to the backaction qubit relaxation events have on the cavity, this errors can propagate and make subsequent bits partially loss the information they had on the original displacement.
    
    \subsection{Backaction evading protocol}

    \begin{figure}[t]
        \centering
        \includegraphics[width=\linewidth]{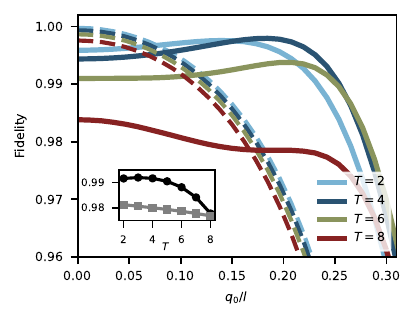}
        \caption{Averaged recovery fidelities as a function of the initial displacement $q_0$ in the presence of noise, with Bayesian estimation and Gaussian prior with standard deviation $\sigma=0.15l$ for different $T$, $M$ values of the backaction evading protocol with $T+M=12$. Solid (dashed) lines are the achieved fidelity with (without) the recovery displacement. We use an envelope with $\Delta=0.3$. The dark (light) markers in the inset show the weighted by the prior $p(q_0)=\mathcal G_\sigma(q_0)$ averaged recovery fidelity with (without) the recovery displacement.}
        \label{fig:noisy-BA-fidelities}
    \end{figure}

    To test the backaction evading protocol in presence of noise, we first compute the average recovery fidelities, as done in \cref{appendix:backaction}, but now starting from the steady state of the sBs protocol in presence of noise $\rho_{\#_\Delta,\eta}$. We plot these fidelities (solid lines) for different values of $T$ in \cref{fig:noisy-BA-fidelities}, with noise strength set by $\eta=1.0$, total bit budget of $T+M=12$, and standard deviation of the Gaussian prior $\sigma=0.15l$. We also plot the achieved fidelities without the recovery displacement (dashed lines). Contrary to the noiseless case, the fidelities without recovery no longer are the same, compare with \cref{fig:avg-recovery-fidelity}. This shows that measuring, instead of resetting the qubit, makes the propagation of qubit errors onto the cavity more pronounced. In the inset we plot the weighted over the prior averaged recovery fidelity. For increasing $T$, these fidelities decrease, both with and without recovery. Hence, to improve the performance the optimization of the recovery function is crucial. Notably, for $T\in [2,6]$ we find that the weighted over the prior averaged fidelities are larger than those obtained in the noiseless setting, both with and without recovery.

    \begin{figure}
        \centering
        \includegraphics[width=\linewidth]{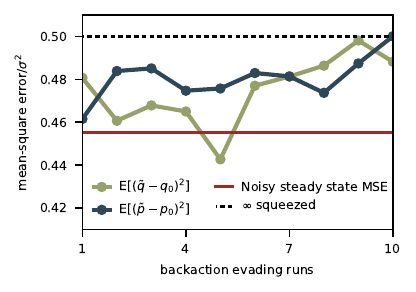}
        \caption{Backaction evading performance in the presence of noise. The noise strength is set by $\eta=1.0$. Blue lines are the mean-squared error obtained running the backaction evading protocol with $T=M=6$, $N=10$, repeated 4000 times. Red solid line is the mean-square error obtained when 6 bits per quadrature are acquired, and the finite-energy sBs steady state in the presence of noise $\rho_{\#_\Delta,\eta}$ is used as sensor state. Dashed black line is the Gaussian limit of displacement estimation. Dotted-dashed lines, added as guide to the eyes, correspond to the average of the mean-square errors from the second run forward. The envelope is set with $\Delta=0.3$. }
        \label{fig:noisy-BA-performance}
    \end{figure}

    We repeat the backaction evading protocol done in the main text, but now in the presence of noise setting $\eta=1.0$. In \cref{fig:noisy-BA-performance} we plot the mean-square error in the estimation of the $q$ and $p$ quadrature displacements with $N=10$, $T=M=6$, averaged over 4000 samples of the whole sequence. Here, we choose $T=6$ instead of $T=8$ due to the sharp decrease in the weighted over the prior averaged fidelity. Also, smaller values of $T$ do not allow to consistently beat the Gaussian limit of sensing, making $T=6$ the ideal choice here. As benchmarks, we also plotted the mean-square error obtained in the estimation of the $q$ quadrature when 6 bits per quadrature are gathered, and the steady state of the sBs protocol $\rho_{\#_\Delta,\eta}$ is used as sensor state. The backaction evading performance approaches this one, showing that the sBs metrology protocol can function as a backaction evading sensor in the presence of noise. Finally, we note that the fluctuations in performance are due to the variance of the mean-square error, as discussed in \cref{appendix:metrological-performance}.

    \newpage

    \bibliography{bibliography.bib}

\end{document}

%% file: commands.tex
\newcommand{\vphi}{\vphi}

\newcommand{\Ev}{\text{E}}
\newcommand{\Var}{\text{Var}}

\usepackage[dvipsnames]{xcolor}